\documentclass[preprint,pre,showpacs]{revtex4}
\usepackage{amscd}
\usepackage{amsfonts}
\usepackage{amsmath}
\usepackage{amssymb}
\usepackage{undertilde}
\usepackage{color}
\usepackage{graphicx}
\usepackage{graphics}
\usepackage{wrapfig}
\usepackage[T1]{fontenc}
\usepackage[latin1]{inputenc}
\usepackage{gensymb}
\def\>{\rangle}
\def\<{\langle}
\def\n{\nonumber}

\newcommand{\id}{\openone}
\begin{document}
\title{Quantal-classical fluctuation relation and the second law of thermodynamics: The quantum linear oscillator}
\author{Ilki Kim}
\email{hannibal.ikim@gmail.com} \affiliation{Joint School of
Nanoscience and Nanoengineering, North Carolina A$\&$T State
University, Greensboro, NC 27411}
\date{\today}
\begin{abstract}
In this work, we study the fluctuation relation and the second law
of thermodynamics within a quantum linear oscillator externally
driven over the period of time $t = \tau$. To go beyond the standard
approach (the two-point projective measurement one) to this subject
and also render it discussed in both quantum and classical domains
on the single footing, we recast this standard approach in terms of
the Wigner function and its propagator in the phase space $(x,p)$.
With the help of the canonical transformation from $(x,p)$ to the
angle-action coordinates $(\phi,\mathbb{I})$, we can then derive a
measurement-free (classical-like) form of the Crooks fluctuation
relation in the Wigner representation. This enables us to introduce
the work $\mathbb{W}_{(\scriptscriptstyle
\mathbb{I}_0,\mathbb{I}_{\tau})}$ associated with a single run from
$(\mathbb{I}_0)$ to $(\mathbb{I}_{\tau})$ over the period $\tau$,
which is a quantum generalization of the thermodynamic work with its
roots in the classical thermodynamics. This quantum work differs
from the energy difference $e_{(\scriptscriptstyle
\mathbb{I}_0,\mathbb{I}_{\tau})} = e(\mathbb{I}_{\tau}) -
e(\mathbb{I}_0)$ unless $\beta,\hbar\to 0$. Consequently, we will
obtain the quantum second-law inequality $\Delta F_{\beta} \leq
\<{\mathbb W}\>_{\scriptscriptstyle \mathbb{P}} \leq
\<e\>_{\scriptscriptstyle \mathbb{P}} = \Delta U$, where ${\mathbb
P}, \Delta F_{\beta}$, and $\<{\mathbb W}\>_{\scriptscriptstyle
\mathbb{P}}$ denote the work (quasi)-probability distribution, the
free energy difference, and the average work distinguished from the
internal energy difference $\Delta U$, respectively, while
$\<{\mathbb W}\>_{\scriptscriptstyle \mathbb{P}} \to \Delta U$ in
the limit of $\beta,\hbar \to 0$ only. Therefore, we can also
introduce the quantum heat $\mathbb{Q}_{\mbox{\scriptsize q}} =
\Delta U - \mathbb{W}$ even for a thermally isolated system,
resulting from the quantum fluctuation therein. This is a more
fine-grained result than $\<{\mathbb W}\>_{\scriptscriptstyle
\mathbb{P}} \equiv \Delta U$ obtained from the standard approach.
Owing to the measurement-free nature of the thermodynamic work
$\mathbb{W}_{(\scriptscriptstyle \mathbb{I}_0,\mathbb{I}_{\tau})}$,
our result can also apply to the (non-thermal) initial states
$\hat{\rho}_0 = (1-\gamma)\,\hat{\rho}_{\beta} +
\gamma\,\hat{\sigma}$ with $\hat{\sigma} \ne \hat{\rho}_{\beta}$.
\end{abstract}
\pacs{03.65.Ta, 11.10.Lm, 05.45.-a}
\maketitle
%
%%%%%%%%%%%%%%%%%%%%%%%%%%%%%%%%%%%%%%%%%%%%%%%%%%%%%%%%%%%%%%%%%%%%%%%%%%%%%
\section{Introduction}\label{sec:introduction}
%%%%%%%%%%%%%%%%%%%%%%%%%%%%%%%%%%%%%%%%%%%%%%%%%%%%%%%%%%%%%%%%%%%%%%%%%%%%%
%
Fluctuation relations such as Jarzynski's equality and Crooks'
theorem have attracted a great deal of interest owing to their
nature of the link between non-equilibrium fluctuations and thermal
equilibrium properties of small systems (either classical or
quantal) \cite{JAC97,CRO98}. Here, the stochastic nature of the
thermodynamic work performed on a given system emerges through
(infinitely) many runs of the external driving of the system; in the
classical case, this nature is associated solely with a random
sampling of individual microstates from the initially prepared
(canonically thermal) state of the system, because after such a
sampling, the system becomes thermally isolated and evolves
deterministically under Hamilton's equations. In the quantum case,
on the other hand, such a stochastic nature is associated not only
with the random sampling from the thermal initial state (as a source
of the thermal fluctuation) but also with the quantum fluctuation
existing even during the external driving. In fact, an appropriate
determination of the work and its probability distribution
associated with both thermal and quantum fluctuations, required for
a legitimate form of the quantum fluctuation relation, has been one
of the central issues in the field of quantum thermodynamics.

The standard approach to the quantum fluctuation relation has been
made in the so-called two-point projective measurement (TPM)
framework
\cite{TAL07,KUR00,TAS00,MUK03,HAE08,LUT08,ESP09,CAM11,HAN15,JAR15}:
An isolated quantum system is initially prepared in the thermal
state $\hat{\rho}_0 = \hat{\rho}_{\beta}$ (with $\beta =
1/k_{\mbox{\tiny B}} T$) and then undergoes an external driving
(denoted by a time-dependent Hamiltonian parameter $\lambda_t$). The
probability distribution of the single-run work ($\mathrm w$) for
the system in a forward process of the external driving is then
given by
\begin{equation}\label{eq:prob-density1f}
    P^{(\mbox{\scriptsize f})}(\mathrm w) = \sum_{n,m} \delta({\mathrm w} - \Delta e_{nm})\times P^{(\mbox{\scriptsize f})}_{nm}\,,
\end{equation}
where the energy-eigenvalue difference $\Delta e_{nm} = e_{m(\tau)}
- e_{n(0)}$, between the two outcomes $e_{n(0)}$ and $e_{m(\tau)}$
found from the initial ($t=0$) and final ($t=\tau$) measurements,
and its probability $P^{(\mbox{\scriptsize f})}_{nm} =
P[m(\tau)|n(0)]\,P[n(0)]$ consisting of both initial probability
$P[n(0)] = e^{-\beta e_{n(0)}}/Z_{\beta}(\lambda_0)$ of the $n$th
energy eigenstate, with the partition function
$Z_{\beta}(\lambda_0)$ for the initial state
$\hat{\rho}_{\beta}(\lambda_0)$, and the conditional probability
$P[m(\tau)|n(0)] = |\<m(\tau)|\hat{\mathcal U}|n(0)\>|^2$ for the
run $|n(0)\> \to |m(\tau)\>$ with the unitary operator
$\hat{\mathcal U}$. Here, the probabilistic nature of finding those
two measurement outcomes gives rise to the stochastic nature of the
work. As such, the single-run work is given by ${\mathrm w}_{nm}
\equiv \Delta e_{nm}$.

Likewise, the work probability distribution for the system in a
backward process starting from $\hat{\rho}_{\beta}(\lambda_{\tau})$
is given by
\begin{equation}\label{eq:prob-density1b}
    P^{(\mbox{\scriptsize b})}(\mathrm w) = \sum_{n,m} \delta({\mathrm w} + {\mathrm w}_{nm})\times P^{(\mbox{\scriptsize b})}_{mn}\,,
\end{equation}
where $P^{(\mbox{\scriptsize b})}_{mn} =
P[n(0)|m(\tau)]\,P[m(\tau)]$ with $P[m(\tau)] = e^{-\beta
e_{m(\tau)}}/Z_{\beta}(\lambda_{\tau})$ and $P[n(0)|m(\tau)] =
P[m(\tau)|n(0)]$. It is then straightforward to obtain the Crooks
fluctuation theorem
\begin{equation}\label{eq:crook1}
    P^{(\mbox{\scriptsize f})}_{nm} = P^{(\mbox{\scriptsize b})}_{mn}\; \exp\{\beta\,({\mathrm w}_{nm} - \Delta F_{\beta}^{(\mbox{\scriptsize f})})\}\,,
\end{equation}
in which the free energy difference is $\Delta
F_{\beta}^{(\mbox{\scriptsize f})}(\lambda_{\tau}) =
F_{\beta}(\lambda_{\tau}) - F_{\beta}(\lambda_0)$. With the help of
Eqs. (\ref{eq:prob-density1f}) and (\ref{eq:prob-density1b}), this
will result in the quantum Jarzynski equality in its known form
\begin{equation}\label{eq:jarzynski-tpm0}
    \<e^{-\beta {\mathrm w}}\>_{\scriptscriptstyle{P^{(\mbox{\tiny f})}}} =
    \int d{\mathrm w}\; e^{-\beta {\mathrm w}}\; P^{(\mbox{\scriptsize f})}({\mathrm w}) = e^{-\beta \Delta F_{\beta}^{(\mbox{\tiny f})}}\,.
\end{equation}
Therefore, the free energy difference and the non-equilibrium
fluctuating work can be exactly linked. With the help of the Jensen
inequality, this Jarzynski equality gives rise to
\begin{equation}\label{eq:scond-law-TPM0}
    \Delta F_{\beta}^{(\mbox{\scriptsize
    f})}(\lambda_{\tau}) \leq \<{\mathrm w}\>_{\scriptscriptstyle{P^{(\mbox{\tiny
    f})}}} \equiv \Delta U(\tau)
\end{equation}
as an expression of the second law of thermodynamics in the quantum
domain [cf. (\ref{eq:finer-form-0})]. By construction, the
non-equilibrium average work $\<{\mathrm
w}\>_{\scriptscriptstyle{P^{(\mbox{\tiny f})}}}$ is identically
equal to the internal energy difference $\Delta U(\tau) =
\<\hat{H}(\lambda_\tau)\>_{\scriptstyle{\rho_{\tau}}} -
\<\hat{H}(\lambda_0)\>_{\scriptstyle{\rho_0}}$ between the initial
and final instants of time, where
$\<\hat{H}(\lambda_t)\>_{\scriptstyle{\rho_t}} =
\mbox{Tr}\{\hat{H}(\lambda_t)\,\hat{\rho}(t)\}$.

In spite of its great usefulness, the TPM framework has a conceptual
issue when it comes to its generalization: It is nonlegitimate to
apply the same form of the work probability to the processes
starting from the non-thermal states $\hat{\rho}_0 \ne
\hat{\rho}_{\beta}$ with coherence in the energy basis; because the
initial projective measurement then destroys the initial coherence
and so produces extra entropy, thus leading to disturbing the
original time evolution of the system. As such, this standard
approach is not fully quantum-mechanical. Moreover, we also note
that an individual external driving ($\lambda_t$) itself, described
by unitary dynamics, produces no entropy at all, regardless of the
initial outcomes $e_{n(0)}$; but an appearance of the entropy
production, achieved through the (classical) mixture over many runs,
is due to the (non-unitary) final projective measurement (thus
viewed as an extra non-equilibrium work). Consequently, it still
remains an open question to introduce a generalized form of quantum
work legitimate for the non-thermal initial condition and the
external driving only.

To go beyond such a limitation, we intend to introduce in this paper
an alternative definition of the quantum work ($\ne {\mathrm w}$)
and its distribution formulated without the projective measurements.
For this purpose, we will resort to the classical phase space
$(x,p)$, in which both quantum and classical fluctuation relations
can be discussed on the single footing, for example, by making use
of the Wigner function and its propagator [cf. Eqs.
(\ref{eq:bopp_0-1})-(\ref{eq:wigner-propagator2})]. In fact, the
Wigner representation is known to be the most classical-like in
propagation among different phase-space representations
\cite{SEG01}. Then we will formulate a Crooks fluctuation theorem in
its measurement-free (classical-like) form by recasting the
characteristic function of the TPM framework in the Wigner
representation (cf. e.g., \cite{DEF13} for a different phase-space
approach without such a link with the TPM approach). This will
finally give a new definition of the work $\mathbb{W}$ as the direct
quantum counterpart to the thermodynamic work which has its original
roots in the classical thermodynamics. For this formulation, we will
also employ the angle-action coordinates $(\phi,{\mathbb I})$ [cf.
Eq. (\ref{eq:action-1})]; as is well-known, this pair is
well-defined for the separable systems (e.g., the generic
one-dimensional ones) and useful for the semiclassical analysis
\cite{BRA97,LAH98,MIL74,MIL16,DIT16}. To our best knowledge, these
coordinates have not been applied extensively for the study of
quantum fluctuation relations. For the sake of an explicit treatment
with analytical rigor, we will restrict our analysis here to a
linear oscillator with its time-dependent frequency $\lambda_t =
\omega(t)$ (cf. \cite{LUT08} for an analysis of this system in the
TPM framework); our methodology will also apply to a more generic
class of systems.

Further, as is well-known, the Wigner function can be
negative-valued, which reflects the quantum fluctuation. This will
result in the work distribution ${\mathbb P}(\mathbb{W})$ with its
negativity. Therefore, our concern lies in the average values only
(over many runs) that have the physical meaning (cf. the unavoidable
negativity of the work distribution has already been studied in the
extended TPM framework where the initial state is non-diagonal in
the energy basis, e.g., \cite{ALL14,MIL17}). Then, it will be shown
that the average work $\<{\mathbb W}\>_{\scriptscriptstyle
\mathbb{P}}$ is distinguished from the internal energy difference
$\Delta U$ which remains unaffected under this transformation of the
representation. For comparison, we also point out that it is
impossible to consider the same scenario (free from the projective
measurements) for such an alternative definition of the classical
work, because in the classical setup, no projective measurements are
required anyway. Therefore, our alternative approach to the quantum
work ($\<{\mathbb W}\>_{\scriptscriptstyle \mathbb{P}}$) will
produce, in the classical limit, no difference from the TPM approach
($\Delta U$). This will also enable us to introduce the quantum heat
$\mathbb{Q}_{\mbox{\scriptsize q}} = \Delta U - \<{\mathbb
W}\>_{\scriptscriptstyle \mathbb{P}} \geq 0$ (with no classical
counterpart) for a thermally isolated system such that
$\mathbb{Q}_{\mbox{\scriptsize d}} + \mathbb{Q}_{\mbox{\scriptsize
q}} = \Delta U - \Delta F_{\beta}$; here, the thermal heat
$\mathbb{Q}_{\mbox{\scriptsize d}} = \<{\mathbb
W}\>_{\scriptscriptstyle \mathbb{P}} - \Delta F_{\beta} \geq 0$
corresponds to the dissipative heat which will go out to the
environment until the system-environment equilibrium will be
achieved, if additional heat exchange between system and environment
is carried out after completing the external driving. Consequently,
we will acquire, as one of our main findings, the quantum second-law
inequality (\ref{eq:finer-form-0}) in a more fine-grained form than
the inequality (\ref{eq:scond-law-TPM0}). Owing to the
measurement-free nature of $\mathbb{W}$ and ${\mathbb
P}(\mathbb{W})$, our result for the thermodynamic work in the
quantum regime will be further generalized to the initial states
being partially thermal in the form of $\hat{\rho}_0 =
(1-\gamma)\,\hat{\rho}_{\beta} + \gamma\,\hat{\sigma}$ with
$\hat{\sigma} \ne \hat{\rho}_{\beta}$.

The general layout of this paper is as follows: In Sec.
\ref{sec:wigner} we provide the phase-space formulation needed for
an introduction of our quantum work and its distribution. In Sec.
\ref{sec:second} we derive the quantal-classical Crooks fluctuation
theorem in our framework and then discuss the second law of
thermodynamics and its implications. In Sec. \ref{sec:non-thermal}
our framework is generalized to the partially thermal initial states
and then several examples of those initial states (with coherence in
the energy basis) are explicitly considered. Finally, we provide
concluding remarks in Sec. \ref{sec:conclusion}.

%%%%%%%%%%%%%%%%%%%%%%%%%%%%%%%%%%%%%%%%%%%%%%%%%%%%%%%%%%%%%%%%%%%%%%%%%%%%%
\section{Phase-Space Formulation and Quantum Work Distribution}\label{sec:wigner}
%%%%%%%%%%%%%%%%%%%%%%%%%%%%%%%%%%%%%%%%%%%%%%%%%%%%%%%%%%%%%%%%%%%%%%%%%%%%%
%
%%%%%%%%%%%%%%%%%%%%%%%%%%%%%%%%%%%%%%%%%%%%%%%%%%%%%%%%%%%%%%%%%%%%%%%%%%%%%
\subsection{Wigner function and its Propagator}
%%%%%%%%%%%%%%%%%%%%%%%%%%%%%%%%%%%%%%%%%%%%%%%%%%%%%%%%%%%%%%%%%%%%%%%%%%%%%
%
To take into consideration the phase-space counterpart to the
forward work distribution in Eq. (\ref{eq:prob-density1f}), we will
make use of the Wigner function (representing the initial
distribution corresponding to $P[n(0)]$ therein)
\cite{WIG32,HIL84,LEE95,BUZ95,SCH01,CUR05}
\begin{equation}\label{eq:bopp_0-1}
    W_{\rho}(x,p) = \frac{1}{2\pi\hbar} \int_{-\infty}^{\infty} d{\xi}\,
    \left\<x + \frac{\xi}{2}\right|\hat{\rho}\left|x - \frac{\xi}{2}\right\>\, \exp\left(-\frac{i}{\hbar} p
    \xi\right)
\end{equation}
(for $\hat{\rho} = \hat{\rho}_{\beta}$) and its propagator
(corresponding to the conditional probability $P[m(\tau)|n(0)]$)
\cite{LEA68,GRO72,SMI78}
\begin{eqnarray}\label{eq:wigner-propagator2}
    \hspace*{-.3cm}&&T_{\mbox{\tiny{W}}}^{(\mbox{\scriptsize f})}(x,p;\tau|x',p';0) := (2\pi\hbar)^{-1}\,
    \mbox{Tr}\{\hat{\Delta}(x,p)\,\hat{\mathcal U}(\tau)\,\hat{\Delta}(x',p')\,\hat{\mathcal U}^{\dagger}(\tau)\}\, =\\
    \hspace*{-.3cm}&&(2\pi\hbar)^{-1}\int d\xi d\xi'\,\exp\left[-\frac{i}{\hbar}(p\xi + p'\xi')\right]\,
    K\left(x+\frac{\xi}{2};\tau\right|\left.x'-\frac{\xi'}{2};0\right)\,
    K^{\ast}\left(x-\frac{\xi}{2};\tau\right|\left.x'+\frac{\xi'}{2};0\right)\,,\n
\end{eqnarray}
in which the operator $\hat{\Delta}(x,p) = \int_{-\infty}^{\infty}
d\xi\,|x-\xi/2\>\<x+\xi/2|\, e^{-i p \xi/\hbar}$ and the usual
propagator $K(x;\tau|x';0) = \<x|\hat{\mathcal U}(\tau)|x'\>$; e.g.,
for the sudden switch ($\hat{\mathcal U}_{\mbox{\scriptsize s}} =
\id$), we can easily obtain
\begin{equation}\label{eq:sudden-switch}
    T_{\mbox{\tiny{W},\scriptsize{s}}}^{(\mbox{\scriptsize f})}(x,p;\tau|x',p';0) =
    \exp\left\{\frac{2i}{\hbar}\,(x-x')\,p\right\}\, \delta(x-x')\, \delta(p-p')\,.
\end{equation}

As is well-known, the Wigner function satisfies its marginal
probability distributions
\begin{equation}\label{eq:marginal-properties0}
    \int dp\; W_{\rho}(x,p) = \<x|\hat{\rho}|x\> \;\;\; ; \;\;\; \int dx\; W_{\rho}(x,p) =
    \<p|\hat{\rho}|p\>
\end{equation}
and gives the expectation value
\begin{eqnarray}\label{eq:bopp_3}
    \mbox{Tr}(\hat{\rho}\,\hat{A})\, =\, \int dx \int dp\; W_{\rho}(x,p)\; A(x,p)
\end{eqnarray}
together with the Weyl-Wigner $c$-number representation of the
observable $\hat{A}$ given by
\begin{subequations}
\begin{eqnarray}
    A(x,p) &=& \int_{-\infty}^{\infty} d{\xi}\,
    \exp\left(-\frac{i}{\hbar} p\,\xi\right)\, \left\<x + \frac{\xi}{2}\right|\hat{A}\left|x -
    \frac{\xi}{2}\right\>\label{eq:A_from wigner0}\\
    \hat{A} &=& (2\pi\hbar)^{-1} \int d\xi \int dx\,dp\, \left|x + \frac{\xi}{2}\right\>\,
    A(x,p)\, \exp\left(\frac{i}{\hbar}
    p\,\xi\right) \left\<x - \frac{\xi}{2}\right|\,.\label{eq:A_from wigner}
\end{eqnarray}
\end{subequations}
Similarly, we have
\begin{equation}\label{eq:rho_sqr0}
    \mbox{Tr}(\hat{\rho}_1\,\hat{\rho}_2) = 2\pi\hbar \int dx \int dp\;
    W_{\rho_1}(x,p)\; W_{\rho_2}(x,p)\,.
\end{equation}
And it is the propagator $T_{\mbox{\tiny{W}}}^{(\mbox{\scriptsize
f})}(x,p;\tau|x',p';0)$ that generates the trajectory running from
the position $(x',p')$ at $t=0$ to $(x,p)$ at $t=\tau$. In the limit
of $\hbar \to 0$, those Wigner trajectories exactly reduce to the
classical trajectories. It is also easy to verify that
\begin{equation}\label{eq:wigner-propagator-id0}
    \int dx' dp'\, T_{\mbox{\tiny{W}}}^{(\mbox{\scriptsize
    f})}(x,p;\tau|x',p';0) = \int dx dp\, T_{\mbox{\tiny{W}}}^{(\mbox{\scriptsize
    f})}(x,p;\tau|x',p';0) = 1
\end{equation}
and
\begin{subequations}
\begin{eqnarray}
    \int dp dp'\, T_{\mbox{\tiny{W}}}^{(\mbox{\scriptsize f})}(x,p;\tau|x',p';0) &=& (2\pi\hbar)\, |K(x;\tau|x';0)|^2 \geq 0\label{eq:wigner-propagator1-1}\\
    \int dx dx'\, T_{\mbox{\tiny{W}}}^{(\mbox{\scriptsize f})}(x,p;\tau|x',p';0) &=& (2\pi\hbar)\, |\tilde{K}(p;\tau|p';0)|^2 \geq 0\,,\label{eq:wigner-propagator1-2}
\end{eqnarray}
\end{subequations}
where $\tilde{K}(p;\tau|p';0) = \<p|\hat{\mathcal U}(\tau)|p'\>$.
The time-evolution $\hat{\rho}_{\tau} = \hat{\mathcal
U}(\tau)\,\hat{\rho}_0\,\hat{\mathcal U}^{\dagger}(\tau)$ is then
rewritten in the Wigner representation as
\begin{equation}\label{eq:wigner-propagator1}
    W_{\scriptstyle{\rho_{\tau}}}(x,p) = \int dx' dp'\, T_{\mbox{\tiny{W}}}^{(\mbox{\scriptsize f})}(x,p;\tau|x',p';0)\,
    W_{\scriptstyle{\rho_0}}(x',p')\,.
\end{equation}
As such, a given final position $(x,p)$ is associated with all
possible initial positions $(x',p')$ through infinitely many
trajectories. Likewise, the Wigner propagator for the backward
process [cf. Eq. (\ref{eq:prob-density1b})] is given by
$T_{\mbox{\tiny{W}}}^{(\mbox{\scriptsize b})}(x',p';\tau|x,p;0) :=
(2\pi\hbar)^{-1} \mbox{Tr}\{\hat{\Delta}(x',p')\,\hat{\mathcal
U}^{\dagger}(\tau)\,\hat{\Delta}(x,p)\,\hat{\mathcal U}(\tau)\}$,
which is easily shown to be identical to
$T_{\mbox{\tiny{W}}}^{(\mbox{\scriptsize f})}(x,p;\tau|x',p';0)$.

For a quantum linear oscillator with a time-dependent frequency
$\omega(t)$, we have the propagator in the Gaussian form
\cite{HUS53}
\begin{equation}\label{eq:propagator-osc1}
    K(x;t|x';0) = \left(\frac{m}{2\pi i\hbar X}\right)^{1/2}
    \exp\left[\frac{i m}{2\hbar X} \left\{x^2 \dot{X} - 2 x x' + (x')^2\,Y\right\}\right]\,,
\end{equation}
where both quantities $X = X(t)$ and $Y = Y(t)$, with $(X(0) = 0,
\dot{X}(0) = 1)$ and $(Y(0) = 1, \dot{Y}(0) = 0)$, are the solutions
to the classical equation of motion $\ddot{X} + \{\omega(t)\}^2\,X =
0$. Then it is straightforward that with the help of Eq.
(\ref{eq:propagator-osc1}), the Wigner propagator in
(\ref{eq:wigner-propagator2}) will be evaluated explicitly.

%%%%%%%%%%%%%%%%%%%%%%%%%%%%%%%%%%%%%%%%%%%%%%%%%%%%%%%%%%%%%%%%%%%%%%%%%%%%%
\subsection{Quantum Work Distribution and Angle-Action Coordinates}
%%%%%%%%%%%%%%%%%%%%%%%%%%%%%%%%%%%%%%%%%%%%%%%%%%%%%%%%%%%%%%%%%%%%%%%%%%%%%
%
In the classical scenario, on the other hand, the work distribution
for a thermally isolated system in a forward process starting from
the thermal state can be expressed as \cite{JAR15}
\begin{equation}\label{eq:prob-density1-classical-f}
    (P_{\mbox{\scriptsize c}})^{(\mbox{\scriptsize f})}(w) = \int\int dE_{\tau}\,dE_0\; \delta(w - W_{0\tau})\times (P_{\mbox{\scriptsize c}})^{(\mbox{\scriptsize f})}_{0\tau}\,,
\end{equation}
in which the energy difference $W_{0\tau} = E_{\tau} - E_0$ as a
single-run work, and its probability density $(P_{\mbox{\scriptsize
c}})^{(\mbox{\scriptsize f})}_{0\tau} =
P_c(E_{\tau}|E_0)\,P_c(E_0)$; here, we have the instantaneous energy
$E_t = H(z_t;\lambda_t)$ with its trajectory $z_t = (x_t,p_t)$
evolving from $z_0$ under Hamilton's dynamics, and the initial
probability density $P_{\mbox{\scriptsize c}}(E_0) = \{e^{-\beta
E_0}/Z_{\beta,\mbox{\scriptsize c}}(\lambda_0)\}\, g(E_0)$ with the
classical partition function $Z_{\beta,\mbox{\scriptsize
c}}(\lambda_0) = e^{-\beta F_{{\scriptscriptstyle \beta},\mbox{\tiny
c}}(\lambda_0)}$ and the density of states $g(E_0)$, and the
conditional probability density $P_c(E_{\tau}|E_0)$ for $E_0 \to
E_{\tau}$. As such, this expression of the classical work
distribution directly shows the formal similarity to the
quantum-mechanical result in Eq. (\ref{eq:prob-density1f}).

Motivated by such an analogy between the quantum and classical work
distributions, we begin by rewriting the Fourier transform of Eq.
(\ref{eq:prob-density1f}) \cite{TAL07}
\begin{subequations}
\begin{equation}\label{eq:hilbert-pic1}
    A^{(\mbox{\scriptsize f})}(u) := \int d{\mathrm w}\, P^{(\mbox{\scriptsize f})}({\mathrm w})\; e^{i u {\mathrm w}} = \sum_{n,m} P[m(\tau)|n(0)]\, P[n(0)]\, e^{i u \{e_{m(\tau)} - e_{n(0)}\}}
\end{equation}
into its phase-space counterpart: By using Eqs.
(\ref{eq:bopp_0-1})-(\ref{eq:wigner-propagator2}) with
(\ref{eq:rho_sqr0}) and (\ref{eq:wigner-propagator1}), we can
acquire
\begin{equation}\label{eq:phase-space1}
    A^{(\mbox{\scriptsize f})}(u) = \int dx dp \int dx' dp'\, \Xi_u^{(\mbox{\scriptsize
    f})}(x',p';x,p)\, T_{\mbox{\tiny{W}}}^{(\mbox{\scriptsize f})}(x,p;\tau|x',p';0)\,,
\end{equation}
\end{subequations}
where the factor
\begin{equation}\label{eq:phase-space1-00}
    \Xi_u^{(\mbox{\scriptsize f})}(x',p';x,p) = \frac{2\pi\hbar}{Z_{\beta}(\lambda_0)} \sum_{n,m}
    W_{n(0)}(x',p';\lambda_0)\, e^{-(\beta + i u) e_{n(0)}}\, W_{m(\tau)}(x,p;\lambda_{\tau})\, e^{i u
    e_{m(\tau)}}\,.
\end{equation}
With the help of the relation $\sum_m W_m(x,p) = (2\pi\hbar)^{-1}$,
it is easy to observe here that
\begin{equation}\label{eq:identities_0}
    \Xi_0^{(\mbox{\scriptsize f})}(x',p';x,p) =
    W_{\beta}(x',p';\lambda_0)\,.
\end{equation}

Now we restrict our discussion to a driven linear oscillator with
$\lambda_t = \omega_t$. To obtain the quantum work distribution in
the Wigner representation taking the form of its classical
counterpart in Eq. (\ref{eq:prob-density1-classical-f}), we employ
the change of coordinates from $(x', p'; x, p)$ to the angle-action
pairs $(\phi_0, \mathbb{I}_0; \phi_{\tau}, \mathbb{I}_{\tau})$
associated with the initial position $(\phi_0, \mathbb{I}_0)$ and
the final position $(\phi_{\tau}, \mathbb{I}_{\tau})$; here
\cite{BRA97,DIT16}
\begin{equation}\label{eq:action-1}
    {\mathbb I} = \frac{1}{2\pi} \oint p\,dx \geq 0\,,
\end{equation}
where the symbol $\oint$ denotes the integral which runs over a
single period in the phase space. For a linear oscillator, it
follows that $x = \{2 \mathbb{I}_t/(m \omega_t)\}^{1/2}\,\sin\phi_t$
and $p = (2 m \omega_t \mathbb{I}_t)^{1/2}\,\cos\phi_t$ with $E_t =
\omega_t\,\mathbb{I}_t$. This enables us to rewrite Eq.
(\ref{eq:phase-space1}) as
\begin{equation}\label{eq:phase-space3}
    A^{(\mbox{\scriptsize f})}(u) = \int_0^{\infty}\int_0^{\infty} d\mathbb{I}_0\, d\mathbb{I}_{\tau}\; \widetilde{\Xi}_u^{(\mbox{\scriptsize
    f})}(\mathbb{I}_0,\mathbb{I}_{\tau}) \times
    B^{(\mbox{\scriptsize f})}(\mathbb{I}_{\tau}|\mathbb{I}_0)\,,
\end{equation}
where the two-state quantity [cf. (\ref{eq:phase-space1-00})] and
the conditional distribution are
\begin{subequations}
\begin{eqnarray}
    \hspace*{-0.5cm}\widetilde{\Xi}_u^{(\mbox{\scriptsize f})}(\mathbb{I}_0,\mathbb{I}_{\tau}) &=&
    \frac{(2\pi\hbar)\, Z_{(\beta \to \beta+iu)}(\omega_0)\, Z_{(\beta \to -iu)}(\omega_{\tau})}{Z_{\beta}(\omega_0)}\, W_{(\beta \to \beta+iu)}(\mathbb{I}_0;\omega_0)\,
    W_{(\beta \to -iu)}(\mathbb{I}_{\tau};\omega_{\tau})\label{eq:phase-space3-2}\\
    \hspace*{-0.5cm}B^{(\mbox{\scriptsize f})}(\mathbb{I}_{\tau}|\mathbb{I}_0) &=& \int_0^{2\pi}\int_0^{2\pi}
    d\phi_{\tau}\, d\phi_0\; \tilde{T}_{\mbox{\tiny{W}}}^{(\mbox{\scriptsize
    f})}(\phi_{\tau},\mathbb{I}_{\tau};\tau|\phi_0,\mathbb{I}_0;0)\,,\label{eq:phase-space3-3}
\end{eqnarray}
\end{subequations}
respectively [cf.
(\ref{eq:wigner-propagator1-1})-(\ref{eq:wigner-propagator1-2})];
the propagator $T_{\mbox{\tiny{W}}}^{(\mbox{\scriptsize
f})}(x,p;\tau|x',p';0) \to
\tilde{T}_{\mbox{\tiny{W}}}^{(\mbox{\scriptsize
f})}(\phi_{\tau},\mathbb{I}_{\tau};\tau|\phi_0,\mathbb{I}_0;0)$.
Here we adopted the Wigner function of the $n$th energy eigenstate
\cite{BUZ95,SCH01}
\begin{equation}\label{eq:wignerfkt-eigen1}
    W_n(x,p) = \frac{(-1)^n}{\pi\hbar}\, e^{-2\,|\eta(x,p)|^2}\;
    L_n(4\,|\eta(x,p)|^2)\,,
\end{equation}
in which the $n$th Laguerre Polynomial $L_n(A)$ and $\eta =
{2}^{-1/2}\,\{\kappa\,x + i p\,(\hbar \kappa)^{-1}\}$ with $\kappa =
(m \omega/\hbar)^{1/2}$, and then applied the identity
$\sum_{n=0}^{\infty} L_n(A)\, z^n = (1-z)^{-1}\, e^{A z/(z-1)}$
\cite{GRA07}, giving rise to the relation $\sum_{n=0} W_n(x,p)\,
e^{-\beta e_n} = Z_{\beta}\, W_{\beta}(x,p)$ indeed; the thermal
Wigner function is given by the Gaussian form \cite{TAN07}
\begin{subequations}
\begin{eqnarray}
    W_{\beta}(x,p;\omega) &=& \frac{\mbox{sech}(\beta\hbar\omega/2)}{(2\pi\hbar)\, Z_{\beta}(\omega)}\,
    \exp\left[-\left(\tanh \frac{\beta\hbar\omega}{2}\right) \left\{(\kappa\,x)^2 + \frac{p^2}{(\hbar
    \kappa)^2}\right\}\right] \geq 0\label{eq:wigner-oscillator1}\\
    \to W_{\beta}(\mathbb{I};\omega) &=& \frac{\mbox{sech}(\beta\hbar\omega/2)}{(2\pi\hbar)\, Z_{\beta}(\omega)}\,
    \exp\left\{-\frac{2\,\mathbb{I}}{\hbar}\,\tanh\left(\frac{\beta\hbar\omega}{2}\right)\right\} \geq 0\,,\label{eq:wigner-oscillator1-0}
\end{eqnarray}
\end{subequations}
where the partition function is given by $Z_{\beta}(\omega) =
2^{-1}\,\mbox{csch}(\beta \hbar\omega/2)$.

Here we take into consideration three particular cases for Eq.
(\ref{eq:phase-space3}); first, the case of $u=0$, in which
$A^{(\mbox{\scriptsize f})}(0) = 1$. Then we can introduce, with the
help of Eq. (\ref{eq:identities_0}), the joint (quasi)probability
distribution associated with a single motion from ($\mathbb{I}_0$)
at $t=0$ to ($\mathbb{I}_{\tau}$) at $t=\tau$
%\begin{subequations}
\begin{equation}\label{eq:proba-forward1}
    {\mathbb P}^{(\mbox{\scriptsize f})}(\mathbb{I}_0,\mathbb{I}_{\tau}) = B^{(\mbox{\scriptsize f})}(\mathbb{I}_{\tau}|\mathbb{I}_0)\; W_{\beta}(\mathbb{I}_0;\omega_0) \geq 0
\end{equation}
with its normalization $\int_0^{\infty}\int_0^{\infty}
d\mathbb{I}_0\, d\mathbb{I}_{\tau}\; {\mathbb P}^{(\mbox{\scriptsize
f})}(\mathbb{I}_0,\mathbb{I}_{\tau}) = 1$; the non-negative nature
of $B^{(\mbox{\scriptsize f})}(\mathbb{I}_{\tau}|\mathbb{I}_0)$ is
verified in Appendix \ref{sec:appendix1}. Likewise, the joint
(quasi)probability distribution for the backward process can also be
acquired
\begin{equation}\label{eq:proba-backward1}
    {\mathbb P}^{(\mbox{\scriptsize b})}(\mathbb{I}_{\tau},\mathbb{I}_0) = B^{(\mbox{\scriptsize b})}(\mathbb{I}_0|\mathbb{I}_{\tau})\; W_{\beta}(\mathbb{I}_{\tau};\omega_{\tau}) \geq
    0\,,
\end{equation}
%\end{subequations}
where $B^{(\mbox{\scriptsize b})}(\mathbb{I}_0|\mathbb{I}_{\tau}) =
B^{(\mbox{\scriptsize f})}(\mathbb{I}_{\tau}|\mathbb{I}_0)$; note
the discussion after Eq. (\ref{eq:wigner-propagator1}). As a result,
we observe that the (quasi)probability distributions ${\mathbb
P}^{(\mbox{\scriptsize f})}(\mathbb{I}_0,\mathbb{I}_{\tau})$ and
${\mathbb P}^{(\mbox{\scriptsize
b})}(\mathbb{I}_{\tau},\mathbb{I}_0)$ are the counterparts to
$P^{(\mbox{\scriptsize f})}_{nm}$ and $P^{(\mbox{\scriptsize
b})}_{mn}$ in Eqs. (\ref{eq:prob-density1f}) and
(\ref{eq:prob-density1b}), respectively.

The second case is given by $-i
\partial_u\,A^{(\mbox{\scriptsize f})}(u)|_{u=0} = \<{\mathrm w}\>_{\scriptscriptstyle{P^{(\mbox{\tiny
f})}}}$ [cf. (\ref{eq:hilbert-pic1})], which equals the internal
energy difference $\Delta U(\tau)$. This quantum-mechanical average
value can now be expressed as
\begin{equation}\label{eq:phase-space-internal1}
    \Delta U(\tau) = \int_0^{\infty} \int_0^{\infty} d\mathbb{I}_0\, d\mathbb{I}_{\tau}\;
    \Delta e_{(\scriptscriptstyle \mathbb{I}_0,\mathbb{I}_{\tau})} \times {\mathbb P}^{(\mbox{\scriptsize f})}(\mathbb{I}_0,\mathbb{I}_{\tau})\,,
\end{equation}
in which the (single-motion) energy difference between the initial
position $\mathbb{I}_0$ and the final position $\mathbb{I}_{\tau}$
is given by
\begin{equation}\label{eq:phase-space9-0}
    \Delta e_{(\scriptscriptstyle \mathbb{I}_0,\mathbb{I}_{\tau})} = (\omega_{\tau}\,\mathbb{I}_{\tau}) -
    (\omega_{0}\,\mathbb{I}_0)\,\{\mbox{sech}(\beta\hbar\omega_{0}/2)\}^2 -
    (\hbar\omega_0/2)\,\{\mbox{tanh}(\beta\hbar\omega_0/2)\}\,.
\end{equation}
To explicitly evaluate Eq. (\ref{eq:phase-space-internal1}), we
obtain the first moment (cf. Appendix \ref{sec:appendix1})
\begin{equation}\label{eq:phase-space10}
    \<\mathbb{I}_{\tau}\>_{{\scriptscriptstyle {\mathbb P}}^{(\mbox{\tiny f})}} = \int_0^{\infty} \int_0^{\infty} d\mathbb{I}_0\, d\mathbb{I}_{\tau}\;
    \mathbb{I}_{\tau} \times {\mathbb P}^{(\mbox{\scriptsize f})}(\mathbb{I}_0,\mathbb{I}_{\tau}) =
    \<\mathbb{I}_0\>_{{\scriptscriptstyle {\mathbb P}}^{(\mbox{\tiny f})}}\; {\mathbb
    K}_{\tau}\,,
\end{equation}
where $\<\mathbb{I}_0\>_{{\scriptscriptstyle {\mathbb
P}}^{(\mbox{\tiny f})}} = (\hbar/2)\, \mbox{coth}(\beta \hbar
\omega_0/2)$ and the (classical) dimensionless quantity
\begin{equation}\label{eq:dimless-1}
    {\mathbb K}_t = \frac{1}{2}\,\left\{\frac{(\dot{X}_t\,Y_t - 1)^2}{\omega_0\,\omega_t\,(X_t)^2} +
    \omega_0\,\omega_t\,(X_t)^2 +
    \frac{\omega_0\,(\dot{X}_t)^2}{\omega_t} +
    \frac{\omega_t\,(Y_t)^2}{\omega_0}\right\}
\end{equation}
with ${\mathbb K}_0 = 1$. By using the inequality of arithmetic and
geometric means, it is easy to see that ${\mathbb K}_t \geq 1$;
e.g., for the sudden switch in Eq. (\ref{eq:sudden-switch}), we find
that ${\mathbb K}_{t,\mbox{\scriptsize{s}}} = (\omega_0/\omega_t +
\omega_t/\omega_0)/2$. If the process is carried out adiabatically,
it turns out that ${\mathbb K}_t = 1$ and so
$\<\mathbb{I}_{\tau}\>_{{\scriptscriptstyle {\mathbb
P}}^{(\mbox{\tiny f})}}$ is invariant. Eqs.
(\ref{eq:phase-space9-0}) and (\ref{eq:phase-space10}) finally give
the internal energy difference, which is, in fact, identical to
\begin{equation}\label{eq:phase-space-12}
    \Delta U(\tau) = \omega_{\tau}\,\<\mathbb{I}_{\tau}\>_{{\scriptscriptstyle {\mathbb P}}^{(\mbox{\tiny f})}} -
    \omega_0\,\<\mathbb{I}_0\>_{{\scriptscriptstyle {\mathbb P}}^{(\mbox{\tiny f})}} =
    \<\mathbb{I}_0\>_{{\scriptscriptstyle {\mathbb P}}^{(\mbox{\tiny f})}}\; ({\omega_{\tau}\,\mathbb K}_{\tau} -
    \omega_0)\,,
\end{equation}
as required [cf. Eqs. (\ref{eq:bopp_3}) and
(\ref{eq:appendix12})-(\ref{eq:appendix14})]. This reduces to
$\beta^{-1}\,\{(\omega_{\tau}/\omega_0)\,{\mathbb K}_{\tau} - 1\}$
in the classical limit.

The third case is given by $(-i
\partial_u)^2\,A^{(\mbox{\scriptsize f})}(u)|_{u=0} = \<{\mathrm w}^2\>_{\scriptscriptstyle{P^{(\mbox{\tiny f})}}} =
\<(e_{m(\tau)} - e_{n(0)})^2\>_{nm}$. This is shown to differ from
$\<(\omega_{\tau}\,\mathbb{I}_{\tau} -
\omega_0\,\mathbb{I}_0)^2\>_{{\scriptscriptstyle {\mathbb
P}}^{(\mbox{\tiny f})}}$, though; in fact, with the help of the
second-moment relations
\begin{subequations}
\begin{eqnarray}
    \<(\mathbb{I}_{\tau})^2\>_{{\scriptscriptstyle {\mathbb P}}^{(\mbox{\tiny f})}} &=& \int_0^{\infty} \int_0^{\infty} d\mathbb{I}_0\, d\mathbb{I}_{\tau}\;
    (\mathbb{I}_{\tau})^2 \times {\mathbb P}^{(\mbox{\scriptsize f})}(\mathbb{I}_0,\mathbb{I}_{\tau}) =
    \{3\,({\mathbb K}_{\tau})^2 - 1\}\; \{\<\mathbb{I}_0\>_{{\scriptscriptstyle {\mathbb P}}^{(\mbox{\tiny f})}}\}^2\label{eq:higher-order-moments-1}\\
    \<(\mathbb{I}_0)^2\>_{{\scriptscriptstyle {\mathbb P}}^{(\mbox{\tiny f})}} &=& \int_0^{\infty} \int_0^{\infty} d\mathbb{I}_0\, d\mathbb{I}_{\tau}\;
    (\mathbb{I}_0)^2 \times {\mathbb P}^{(\mbox{\scriptsize f})}(\mathbb{I}_0,\mathbb{I}_{\tau}) =
    2\; \{\<\mathbb{I}_0\>_{{\scriptscriptstyle {\mathbb P}}^{(\mbox{\tiny f})}}\}^2\label{eq:higher-order-moments-2}\\
    \<\mathbb{I}_0\, \mathbb{I}_{\tau}\>_{{\scriptscriptstyle {\mathbb P}}^{(\mbox{\tiny f})}} &=& \int_0^{\infty} \int_0^{\infty} d\mathbb{I}_0\, d\mathbb{I}_{\tau}\;
    (\mathbb{I}_0\, \mathbb{I}_{\tau}) \times {\mathbb P}^{(\mbox{\scriptsize f})}(\mathbb{I}_0,\mathbb{I}_{\tau}) =
    2\,{\mathbb K}_{\tau}\; \{\<\mathbb{I}_0\>_{{\scriptscriptstyle {\mathbb P}}^{(\mbox{\tiny
    f})}}\}^2\label{eq:higher-order-moments-3}
\end{eqnarray}
\end{subequations}
(cf. Appendix \ref{sec:appendix1}), we can verify that
\begin{subequations}
\begin{eqnarray}\label{eq:difference-bet-tpm-and-mine1}
    \<{\mathrm w}^2\>_{\scriptscriptstyle{P^{(\mbox{\tiny f})}}} &=&
    \<(\omega_{\tau}\,\mathbb{I}_{\tau} - \omega_0\,\mathbb{I}_0)^2\>_{{\scriptscriptstyle {\mathbb P}}^{(\mbox{\tiny
    f})}} + \frac{\hbar^2\,\omega_0\,\omega_{\tau}}{2}\,{\mathbb K}_{\tau} - \frac{\hbar^2}{4}\,\{(\omega_0)^2 +
    (\omega_{\tau})^2\}\\
    &=& \<\{\Delta e_{(\scriptscriptstyle \mathbb{I}_0,\mathbb{I}_{\tau})}\}^2\>_{{\scriptscriptstyle {\mathbb P}}^{(\mbox{\tiny f})}} +
    \left\{\left(\frac{\hbar \omega_0}{2}\right)\,
    \mbox{sech}\left(\frac{\beta\hbar\omega_0}{2}\right)\right\}^2 - \left(\frac{\hbar \omega_{\tau}}{2}\right)^2\,,
\end{eqnarray}
\end{subequations}
in which $\<(\omega_{\tau}\,\mathbb{I}_{\tau} -
\omega_0\,\mathbb{I}_0)^2\>_{{\scriptscriptstyle {\mathbb
P}}^{(\mbox{\tiny f})}} = \{3\,(\omega_{\tau})^2\,({\mathbb
K}_{\tau})^2 - 4\,\omega_0\,\omega_{\tau}\,{\mathbb K}_{\tau} +
2\,(\omega_0)^2 - (\omega_{\tau})^2\}\,
\{\<\mathbb{I}_0\>_{{\scriptscriptstyle {\mathbb P}}^{(\mbox{\tiny
f})}}\}^2 $ [cf. (\ref{eq:phase-space9-0})]. In fact, the
quantum-mechanical expectation value (in the form of the first
moment), such as $\Delta U(\tau)$, is identically evaluated in both
TPM and Wigner frameworks; however, it can be shown that such a
framework-independent behavior is not available any longer for all
higher-order moments $\<(\mathbb{I}_{\tau})^n\,
(\mathbb{I}_0)^m\>_{{\scriptscriptstyle {\mathbb P}}^{(\mbox{\tiny
f})}}$. At this point, we also remind that differing from the energy
operator ($\hat{H}$), the quantum work ($\mathrm w$) performed by an
external agent is not a quantum-mechanical observable \cite{TAL07}.

%%%%%%%%%%%%%%%%%%%%%%%%%%%%%%%%%%%%%%%%%%%%%%%%%%%%%%%%%%%%%%%%%%%%%%%%%%%%%
\section{Quantum Crooks Fluctuation Theorem and the Second Law}\label{sec:second}
%%%%%%%%%%%%%%%%%%%%%%%%%%%%%%%%%%%%%%%%%%%%%%%%%%%%%%%%%%%%%%%%%%%%%%%%%%%%%
%
%%%%%%%%%%%%%%%%%%%%%%%%%%%%%%%%%%%%%%%%%%%%%%%%%%%%%%%%%%%%%%%%%%%%%%%%%%%%%
\subsection{Crooks Theorem in the Wigner Representation}\label{subsec:wigner}
%%%%%%%%%%%%%%%%%%%%%%%%%%%%%%%%%%%%%%%%%%%%%%%%%%%%%%%%%%%%%%%%%%%%%%%%%%%%%
%
We are ready to consider a quantum Crooks fluctuation theorem in the
classical phase space: By combining Eqs. (\ref{eq:proba-forward1})
and (\ref{eq:proba-backward1}), leading to
\begin{equation}\label{eq:crooks-quantum1-0}
    \frac{{\mathbb P}^{(\mbox{\scriptsize
    f})}(\mathbb{I}_0,\mathbb{I}_{\tau})}{{\mathbb P}^{(\mbox{\scriptsize
    b})}(\mathbb{I}_{\tau},\mathbb{I}_0)} = \frac{W_{\beta}(\mathbb{I}_0;\omega_0)}{W_{\beta}(\mathbb{I}_{\tau};\omega_{\tau})}\,,
\end{equation}
we can easily obtain this fluctuation theorem given by
\begin{equation}\label{eq:crooks-quantum1}
    {\mathbb P}^{(\mbox{\scriptsize
    f})}(\mathbb{I}_0,\mathbb{I}_{\tau})
    = {\mathbb P}^{(\mbox{\scriptsize
    b})}(\mathbb{I}_{\tau},\mathbb{I}_0)\; \exp\left\{\beta \left(\mathbb{W}_{(\scriptscriptstyle \mathbb{I}_0,\mathbb{I}_{\tau})} - \Delta F_{\beta}^{(\mbox{\scriptsize
    f})}\right)\right\}\,.
\end{equation}
Here, the free energy difference $\Delta
F_{\beta}^{(\mbox{\scriptsize f})}$ for the forward process is
explicitly given by $F_{\beta}(\omega_{\tau}) - F_{\beta}(\omega_0)
= \beta^{-1}\,
\ln\,[\{\mbox{sinh}(\beta\hbar\omega_{\tau}/2)\}/\{\mbox{sinh}(\beta\hbar\omega_0/2)\}]$,
and the single-motion work associated with the transformation from
the initial position $\mathbb{I}_0$ to the final position
$\mathbb{I}_{\tau}$ is identified as
\begin{eqnarray}\label{eq:phase-space7}
    \mathbb{W}_{(\scriptscriptstyle \mathbb{I}_0,\mathbb{I}_{\tau})}
    &=& -\frac{1}{\beta}\, \ln\left\{\frac{W_{\beta}(\mathbb{I}_{\tau};\omega_{\tau})}{W_{\beta}(\mathbb{I}_0;\omega_0)}\right\} +
    \Delta F_{\beta}^{(\mbox{\scriptsize f})}(\omega_{\tau})\\
    &=& (\omega_{\tau}\,\mathbb{I}_{\tau})\,
    \frac{\mbox{tanh}(\beta\hbar\omega_{\tau}/2)}{\beta\hbar\omega_{\tau}/2} - (\omega_0\,\mathbb{I}_0)\,
    \frac{\mbox{tanh}(\beta\hbar\omega_0/2)}{\beta\hbar\omega_0/2} +
    \frac{1}{\beta}\,
    \ln\left\{\frac{\mbox{cosh}(\beta\hbar\omega_{\tau}/2)}{\mbox{cosh}(\beta\hbar\omega_0/2)}\right\}\,.\n
\end{eqnarray}
By construction, this form of the thermodynamic work in the quantum
regime, linked to $\Delta F_{\beta}^{(\mbox{\scriptsize f})}$, was
derived from requiring the Crooks theorem in the Wigner
representation. Taking Eqs. (\ref{eq:A_from
wigner0})-(\ref{eq:A_from wigner}) into consideration, we find that
this quantum work, expressed in terms of the action coordinates and
formulated without resorting to any projective measurements, is
evidently not a quantum-mechanical observable. Here, we also observe
that this work and $\Delta e_{(\scriptscriptstyle
\mathbb{I}_0,\mathbb{I}_{\tau})}$ in Eq. (\ref{eq:phase-space9-0})
differ from each other, while both become identical ($\omega_{\tau}
\mathbb{I}_{\tau} - \omega_0 \mathbb{I}_0$) in the limit of
$\beta,\hbar \to 0$ where the quantum Crooks theorem in Eq.
(\ref{eq:crooks-quantum1}) reduces to its classical counterpart in
its known form.

Then we can introduce the quantum work distribution for the forward
process in the Wigner representation
%\begin{subequations}
\begin{equation}\label{eq:work-dist-new_1}
    {\mathbb P}^{(\mbox{\scriptsize f})}(\mathbb{W}) = \int_0^{\infty}\int_0^{\infty}
    d\mathbb{I}_0\, d\mathbb{I}_{\tau}\; \delta(\mathbb{W} - \mathbb{W}_{(\scriptscriptstyle \mathbb{I}_0,\mathbb{I}_{\tau})})\; {\mathbb P}^{(\mbox{\scriptsize
    f})}(\mathbb{I}_0,\mathbb{I}_{\tau})
\end{equation}
[cf. Eq. (\ref{eq:prob-density1f})], which is valid in the entire
quantum regime and may also be viewed as the quantum generalization
of the classical work distribution in Eq.
(\ref{eq:prob-density1-classical-f}). Likewise, the work
distribution for the backward process turns out to be
\begin{equation}\label{eq:work-dist-new_2}
    {\mathbb P}^{(\mbox{\scriptsize b})}(\mathbb{W}) = \int_0^{\infty}\int_0^{\infty}
    d\mathbb{I}_{\tau}\, d\mathbb{I}_0\; \delta(\mathbb{W} + \mathbb{W}_{(\scriptscriptstyle \mathbb{I}_0,\mathbb{I}_{\tau})})\;
    {\mathbb P}^{(\mbox{\scriptsize b})}(\mathbb{I}_{\tau},\mathbb{I}_0)
\end{equation}
%\end{subequations}
[cf. Eq. (\ref{eq:prob-density1b})]. With the help of Eqs.
(\ref{eq:work-dist-new_1}) and (\ref{eq:work-dist-new_2}), the
integration of Eq. (\ref{eq:crooks-quantum1}) over $\mathbb{I}_0$
and $\mathbb{I}_{\tau}$ will result in the quantum Jarzynski
equality
\begin{equation}\label{eq:phase-space6}
    \left\<e^{-\beta\,\mathbb{W}}\right\>_{{\scriptscriptstyle {\mathbb P}}^{(\mbox{\tiny f})}} = \int d\mathbb{W}\; e^{-\beta\,\mathbb{W}}\;
    {\mathbb P}^{(\mbox{\scriptsize f})}(\mathbb{W}) = e^{-\beta\,\Delta F_{{\scriptscriptstyle \beta}}^{(\mbox{\tiny
    f})}}\,.
\end{equation}
Employing Eqs. (\ref{eq:phase-space7}) and
(\ref{eq:work-dist-new_1}) with (\ref{eq:phase-space10}), we can
also evaluate the average work such that
\begin{equation}\label{eq:average-work0}
    \left\<\mathbb{W}\right\>_{{\scriptscriptstyle {\mathbb P}}^{(\mbox{\tiny f})}} = \int d\mathbb{W}\; \mathbb{W} \times {\mathbb P}^{(\mbox{\scriptsize
    f})}(\mathbb{W}) = \mathbb{Q}_{\mbox{\scriptsize d}}^{(\mbox{\scriptsize f})} +
    \Delta F_{\beta}^{(\mbox{\scriptsize f})}(\omega_{\tau})\,.
\end{equation}
Here, the dissipative heat is explicitly given by
\begin{equation}\label{eq:dissipative-heat-0}
    \mathbb{Q}_{\mbox{\scriptsize d}}^{(\mbox{\scriptsize f})} = \frac{1}{\beta}\, \left\{\frac{{\mathbb K}_{\tau}}{\widetilde{{\mathbb K}}_{\tau}} - 1
    + \ln(\widetilde{{\mathbb K}}_{\tau})\right\} \geq 0\,,
\end{equation}
where $\widetilde{{\mathbb K}}_{\tau} =
\{\mbox{coth}(\beta\hbar\omega_{\tau}/2)\}/\{\mbox{coth}(\beta\hbar\omega_0/2)\}$;
the minimum value of $\mathbb{Q}_{\mbox{\scriptsize
d}}^{(\mbox{\scriptsize f})}$ is achieved if the entire process is
carried out adiabatically (${\mathbb K}_{\tau} = 1$). With the help
of the inequality given by $y \geq \ln(y) + 1$ (with $y =
1/\widetilde{{\mathbb K}}_{\tau}$), it is then easy to verify that
this minimum value is non-negative, and so the second law of
thermodynamics is met in Eq. (\ref{eq:average-work0}); cf. Eq.
(\ref{eq:average-work0-husimi}) for the same discussion in other
phase-space representations.

Consequently, the average work
$\left\<\mathbb{W}\right\>_{{\scriptscriptstyle {\mathbb
P}}^{(\mbox{\tiny f})}}$ is distinguished from the internal energy
difference in (\ref{eq:phase-space-12}); at zero temperature, we
have $\Delta U(\tau) \to (\hbar/2)\,(\omega_{\tau}\,\mathbb K_{\tau}
- \omega_0)$ but $\left\<\mathbb{W}\right\>_{{\scriptscriptstyle
{\mathbb P}}^{(\mbox{\tiny f})}} = \Delta F_{{\scriptscriptstyle
\beta}}^{(\mbox{\scriptsize f})} \to (\hbar/2)\,(\omega_{\tau} -
\omega_0)$ while in the high-temperature regime ($\beta \to 0$), the
two first moments become identical. Therefore, it is legitimate to
say that this difference should be ascribed to the (non-thermal)
quantum fluctuation. In fact, Figs. \ref{fig:fig1} and
\ref{fig:fig2} show that
\begin{equation}\label{eq:second-law-first-part1}
    \left\<\mathbb{W}\right\>_{{\scriptscriptstyle {\mathbb P}}^{(\mbox{\tiny f})}} \leq \Delta U(\tau) = \<{\mathrm w}\>_{\scriptscriptstyle{P^{(\mbox{\tiny f})}}}\,.
\end{equation}
It is also tempting to examine more rigorously this inequality for
its validity. To do so, we restrict ourselves to the periodic
external drivings ($\omega_\tau = \omega_0$) for arbitrary pairs of
$(\omega_t,\tau)$. Then, it follows from Eqs.
(\ref{eq:phase-space-12}) and (\ref{eq:average-work0}) that the net
external work on the system and the net internal energy difference
are
\begin{equation}\label{eq:phase-space-12-1}
    \left\<\mathbb{W}\right\>_{{\scriptscriptstyle {\mathbb P}}^{(\mbox{\tiny f})},{\mbox{\tiny p}}} = \beta^{-1}\,\{(\mathbb{K}_{\tau})_{\mbox{\tiny p}} -
    1\}\;\;\; ,\;\;\; \Delta U_{\mbox{\tiny p}}^{(\mbox{\scriptsize f})}(\tau) = \left\<\mathbb{W}\right\>_{{\scriptscriptstyle {\mathbb P}}^{(\mbox{\tiny f})},\mbox{\tiny p}}\, y\,
    \mbox{coth}(y)\,,
\end{equation}
respectively, where $y = \beta\hbar\omega_0/2$. Because the factor
$y\, \mbox{coth}(y)$ monotonically increases for $y \geq 0$, it is
easy to see that
\begin{equation}\label{eq:2nd-law-phase1}
    \left\<\mathbb{W}\right\>_{{\scriptscriptstyle {\mathbb P}}^{(\mbox{\tiny
    f})},\mbox{\tiny p}} \leq \Delta U_{\mbox{\tiny p}}^{(\mbox{\scriptsize f})}(\tau)\,.
\end{equation}

Now we are in a position to discuss the quantum second law
associated with $\left\<\mathbb{W}\right\>_{{\scriptscriptstyle
{\mathbb P}}^{(\mbox{\tiny f})}}$: The inequality
(\ref{eq:second-law-first-part1}), together with the Jensen
inequality resulting from Eq. (\ref{eq:phase-space6}), yields the
quantum-thermodynamic inequality
\begin{equation}\label{eq:finer-form-0}
    \Delta F_{{\scriptscriptstyle \beta}}^{(\mbox{\scriptsize
    f})}(\omega_{\tau}) \leq \left\<\mathbb{W}\right\>_{{\scriptscriptstyle {\mathbb P}}^{(\mbox{\tiny
    f})}} \leq \Delta U(\tau)
\end{equation}
as one of our main findings. This represents a more fine-grained
result than the inequality (\ref{eq:scond-law-TPM0}) obtained from
the TPM framework; the first inequality reduces to the equality if
the thermodynamic quasi-static process with
$\mathbb{Q}_{\mbox{\scriptsize d}}^{(\mbox{\scriptsize f})} = 0$
(obtained from both ${\mathbb K}_{\tau} = 1$ and
$\widetilde{{\mathbb K}}_{\tau} = 1$) is carried out (differing from
the adiabatic process with ${\mathbb K}_{\tau} = 1$). This can be
implemented if additional heat exchange between system and
environment is undergone infinitely slowly over the entire process.
On the other hand, the second inequality reduces to the equality in
the limit of $\beta, \hbar \to 0$ only; therefore, we can introduce
the quantum heat $\mathbb{Q}_{\mbox{\scriptsize
q}}^{(\mbox{\scriptsize f})} := \Delta U(\tau) -
\left\<\mathbb{W}\right\>_{{\scriptscriptstyle {\mathbb
P}}^{(\mbox{\tiny f})}} \geq 0$ (even for a thermally isolated
system) which vanishes in the classical limit. This extra heat
$\mathbb{Q}_{\mbox{\scriptsize q}}^{(\mbox{\scriptsize f})}$,
different from the dissipative heat $\mathbb{Q}_{\mbox{\scriptsize
d}}^{(\mbox{\scriptsize f})}$, is accordingly interpreted as a
``built-in'' quantity induced by the (non-thermal) quantum
fluctuation. In Fig. \ref{fig:fig3}, the behaviors of
$\mathbb{Q}_{\mbox{\scriptsize q}}^{(\mbox{\scriptsize f})}$ and
$\mathbb{Q}_{\mbox{\scriptsize d}}^{(\mbox{\scriptsize f})}$ are
explicitly compared (note that $\mathbb{Q}_{\mbox{\scriptsize
q}}^{(\mbox{\scriptsize f})}/\mathbb{Q}_{\mbox{\scriptsize
d}}^{(\mbox{\scriptsize f})} > 1$).

It is also interesting to discuss the difference between two second
moments $\left\<\mathbb{W}^2\right\>_{{\scriptscriptstyle {\mathbb
P}}^{(\mbox{\tiny f})}}$ and $\<{\mathrm
w}^2\>_{\scriptscriptstyle{P^{(\mbox{\tiny f})}}}$ as a next step to
the first-moment inequality (\ref{eq:second-law-first-part1}): With
the help of Eqs.
(\ref{eq:higher-order-moments-1})-(\ref{eq:higher-order-moments-3}),
we can acquire the relative variance
\begin{eqnarray}\label{eq:variance-work1}
    && \left\<(\widetilde{\Delta \mathbb{W}})^2\right\>_{{\scriptscriptstyle {\mathbb P}}^{(\mbox{\tiny f})}}
    =
    \frac{\left\<\mathbb{W}^2\right\>_{{\scriptscriptstyle {\mathbb P}}^{(\mbox{\tiny f})}} -
    \left\{\left\<\mathbb{W}\right\>_{{\scriptscriptstyle {\mathbb P}}^{(\mbox{\tiny f})}}\right\}^2}{\left\<\mathbb{W}^2\right\>_{{\scriptscriptstyle {\mathbb P}}^{(\mbox{\tiny
    f})}}}\n\\
    &=&
    \left[1 + \frac{\beta^2}{4} \left\{\frac{\{2\,({\mathbb K}_{\tau})^2 - 1\}\,
    (\<\mathbb{I}_0\>_{{\scriptscriptstyle {\mathbb P}}^{(\mbox{\tiny f})}}/\hbar)^2}{\{\mbox{coth}(\beta\hbar\omega_{\tau}/2)\}^2} -
    \frac{{\mathbb K}_{\tau}\, \<\mathbb{I}_0\>_{{\scriptscriptstyle {\mathbb P}}^{(\mbox{\tiny f})}}/\hbar}{\mbox{coth}(\beta\hbar\omega_{\tau}/2)} +
    \frac{1}{4}\right\}^{-1} \left\{\left\<\mathbb{W}\right\>_{{\scriptscriptstyle {\mathbb P}}^{(\mbox{\tiny f})}}\right\}^2\right]^{-1}
\end{eqnarray}
which is less than its counterpart $\<(\widetilde{\Delta {\mathrm
w}})^2\>_{\scriptscriptstyle{P^{(\mbox{\tiny f})}}} = 1 - (\Delta
U)^2/\<{\mathrm w}^2\>_{\scriptscriptstyle{P^{(\mbox{\tiny f})}}}$
[cf. Eq. (\ref{eq:difference-bet-tpm-and-mine1})]; at zero
temperature, $\<(\widetilde{\Delta
\mathbb{W}})^2\>_{{\scriptscriptstyle {\mathbb P}}^{(\mbox{\tiny
f})}}$ identically vanishes while $\<(\widetilde{\Delta {\mathrm
w}})^2\>_{\scriptscriptstyle{P^{(\mbox{\tiny f})}}} \to
2\,(\omega_{\tau})^2\,\{({\mathbb K}_{\tau})^2 -
1\}\,\{3\,(\omega_{\tau})^2\,({\mathbb K}_{\tau})^2 -
2\,\omega_0\,\omega_{\tau}\,{\mathbb K}_{\tau} + (\omega_0)^2 -
2\,(\omega_{\tau})^2\}^{-1}$. Again, it is the (non-thermal) quantum
fluctuation contribution to $\<(\widetilde{\Delta {\mathrm
w}})^2\>_{\scriptscriptstyle{P^{(\mbox{\tiny f})}}}$ that gives this
difference. On the other hand, in the high-temperature regime
($\beta \to 0$), these two quantities become identical. Fig.
\ref{fig:fig4} demonstrates this variance difference.

Finally, we point out that the above discussion signifies that
unlike the internal energy difference, the average quantum work (not
viewed as a quantal expectation value) may be contingent upon the
representation in consideration; in fact, there has thus far been no
broadly agreed-upon ``textbook'' definition of quantum work
\cite{JAR15}. Therefore, the choice of an appropriate representation
for the average thermodynamic work in the (entire) quantum regime
with its direct classical counterpart on the same footing will also
be a significant issue.

%%%%%%%%%%%%%%%%%%%%%%%%%%%%%%%%%%%%%%%%%%%%%%%%%%%%%%%%%%%%%%%%%%%%%%%%%%%%%
\subsection{Why only the Wigner Representation for the work distribution?}\label{subsec:wigner-only}
%%%%%%%%%%%%%%%%%%%%%%%%%%%%%%%%%%%%%%%%%%%%%%%%%%%%%%%%%%%%%%%%%%%%%%%%%%%%%
%
In fact, in addition to the Wigner function $W_{\rho}(x,p)$, one has
several other quasi-probability distributions such as the Husimi
function $Q_{\rho}(x,p) \geq 0$, the Glauber-Sudarshan function
$P_{\rho}(x,p)$, the Kirwood function $K_{\rho}(x,p)$, and the
standard-ordered function $F^{(s)}_{\rho}(x,p)$ \cite{LEE95}.
Therefore, it is also tempting to discuss the quantum work
(quasi)probability distribution in these additional representations.
First, we point out that the two functions $K_{\rho}(x,p)$ and
$F^{(s)}_{\rho}(x,p)$ are not always real-valued and therefore will
not be under our consideration here. Therefore, we focus on the two
real-valued functions $Q_{\rho}(x,p)$ and $P_{\rho}(x,p)$ only,
which do not fulfill the marginal-distribution condition, like in
Eq. (\ref{eq:marginal-properties0}) for $W_{\rho}(x,p)$, though. In
fact, the Husimi function can simply be understood as the
convolution of the Wigner function with a Gaussian filter such that
\cite{LEE95}
\begin{equation}\label{eq:husimi_2}
    Q_{\rho}(x,p) = \frac{1}{\pi\hbar} \int dx' dp'\;
    W_{\rho}(x',p')\, \exp\left\{-\left(\{\kappa\,(x'-x)\}^2 +
    \left(\frac{p'-p}{\hbar\kappa}\right)^2\right)\right\}\,.
\end{equation}
Likewise, the relation between the Husimi and Glauber-Sudarshan
functions is
\begin{equation}\label{eq:glauber_2}
    Q_{\rho}(x,p) = \frac{1}{2\pi\hbar} \int dx' dp'\;
    P_{\rho}(x',p')\, \exp\left\{-\frac{1}{2} \left(\{\kappa\,(x'-x)\}^2 +
    \left(\frac{p'-p}{\hbar\kappa}\right)^2\right)\right\}\,.
\end{equation}
As an example, the thermal state for a linear oscillator is given by
the Gaussian form
\begin{eqnarray}
    Q_{\beta}(x,p;\omega_t) &=& \frac{\{\mbox{cosh}(\beta\hbar\omega_t/2) + a_{\scriptscriptstyle{\Upsilon}}\,\mbox{sinh}(\beta\hbar\omega_t/2)\}^{-1}}{(2\pi\hbar)\,
    Z_{\beta}(\omega_t)}\, \times\n\\
    && \exp\left[-\left(\coth \frac{\beta\hbar\omega_t}{2} + a_{\scriptscriptstyle{\Upsilon}}\,\right)^{-1}\, \left\{(\kappa_t\,x)^2 + \frac{p^2}{(\hbar \kappa_t)^2}\right\}\right] \geq 0\label{eq:husimi-oscillator1}
\end{eqnarray}
with $a_{\scriptscriptstyle{\Upsilon}} = 1$, and
$P_{\beta}(x,p;\omega_t)$ obtained from Eq.
(\ref{eq:husimi-oscillator1}) but with
$a_{\scriptscriptstyle{\Upsilon}} = -1$; cf.
$W_{\beta}(x,p;\omega_t)$ with $a_{\scriptscriptstyle{\Upsilon}} =
0$.

The propagators for all other phase-space representations have been
discussed in \cite{SEG01}; in fact, they can be expressed in terms
of the Wigner propagator $T_{\mbox{\tiny{W}}}(x,p;\tau|x',p';0)$ and
the evolution kernels $G$'s such that
\begin{eqnarray}\label{eq:propagators_others0}
    T_{\scriptscriptstyle \Upsilon}(x,p;\tau|x',p';0)
    &=&
    \int dx_1 \int dp_1 \int dx_2 \int dp_2\;
    G_{\mbox{\tiny{W}}\to{\scriptscriptstyle \Upsilon}}(x_1-x,p_1-p)\, \times\n\\
    && T_{\mbox{\tiny{W}}}(x_1,p_1;\tau|x_2,p_2;0)\; G_{{\scriptscriptstyle
    \Upsilon}\to\mbox{\tiny{W}}}(x'-x_2,p'-p_2)\,,
\end{eqnarray}
in which the symbols $\Upsilon = W, Q, P, K, F^{(s)}$ denote the
respective phase-space representations; the kernels are explicitly
given in \cite{SEG01}, e.g., $
G_{\mbox{\tiny{W}}\to\mbox{\tiny{W}}}(x,p) = \delta(x)\,\delta(p)$.
This means that these propagators are equivalent to transforming
first into the Wigner representation $(G_{{\scriptscriptstyle
\Upsilon}\to\mbox{\tiny{W}}}$) and then propagating with
$T_{\mbox{\tiny{W}}}$, followed by transforming into the original
representation ($G_{\mbox{\tiny{W}}\to{\scriptscriptstyle
\Upsilon}}$). Further, it has been shown that free propagation in
the Wigner representation is completely classical-like such that
$T_{\mbox{\tiny{W}}}(x,p;\tau|x',p';0) \to \delta(p-p')\,\delta(x' +
t\,p/m - x)$; however, in all other representations such a simple
one-to-one correspondence between the initial and final phase-space
points is unavailable even for free propagation. This implies that
propagations in all other representations will in general possess
the non-classical features in more complicated form than the Wigner
propagation. Further, unlike $T_{\mbox{\tiny{W}}}$, both propagators
$T_{\mbox{\tiny{Q}}}$ and $T_{\mbox{\tiny{P}}}$ under our
consideration have been shown to involve divergent evolution kernels
$G$'s in Eq. (\ref{eq:propagators_others0}). This divergence for
$\Upsilon = Q, P$ will make it obscure to have the symmetry given by
$T_{\scriptscriptstyle \Upsilon}^{(\mbox{\scriptsize
f})}(x,p;\tau|x',p';0) = T_{\scriptscriptstyle
\Upsilon}^{(\mbox{\scriptsize b})}(x',p';\tau|x,p;0)$ between the
forward and backward processes, which, for $\Upsilon = W$, led to
the result like in Eq. (\ref{eq:crooks-quantum1-0}) and then the
quantal-classical Crooks theorem in its compact form.

Nevertheless, it is still instructive to consider Eq.
(\ref{eq:crooks-quantum1-0}) but now expressed in terms of either
$Q_{\beta}$'s or $P_{\beta}$'s in place of $W_{\beta}$'s. Then, it
will be straightforward to introduce the ``work'' in the Husimi
representation given by
\begin{equation}\label{eq:phase-space-husimi}
    \mathbb{W}_{\mbox{\tiny{Q}},(\scriptscriptstyle \mathbb{I}_0,\mathbb{I}_{\tau})}
    = -\frac{1}{\beta}\, \ln\left\{\frac{Q_{\beta}(\mathbb{I}_{\tau};\omega_{\tau})}{Q_{\beta}(\mathbb{I}_0;\omega_0)}\right\} +
    \Delta F_{\beta}^{(\mbox{\scriptsize f})}(\omega_{\tau})
\end{equation}
($\ne \mathbb{W}_{(\scriptscriptstyle
\mathbb{I}_0,\mathbb{I}_{\tau})}$ in (\ref{eq:phase-space7})) and
likewise the ``work''
$\mathbb{W}_{\mbox{\tiny{P}},(\scriptscriptstyle
\mathbb{I}_0,\mathbb{I}_{\tau})}$ in the Glauber-Sudarshan
representation. By applying the same techniques as for Eqs.
(\ref{eq:average-work0})-(\ref{eq:dissipative-heat-0}), we can
finally arrive at the ``average work''
\begin{equation}\label{eq:average-work0-husimi}
    \left\<\mathbb{W}_{\scriptscriptstyle{\Upsilon}}\right\>_{{\scriptscriptstyle {\mathbb P}}^{(\mbox{\tiny f})}} = \mathbb{Q}_{\scriptscriptstyle{\Upsilon},\mbox{\scriptsize{d}}}^{(\mbox{\scriptsize f})} +
    \Delta F_{\beta}^{(\mbox{\scriptsize f})}(\omega_{\tau})
\end{equation}
($\ne \left\<\mathbb{W}\right\>_{{\scriptscriptstyle {\mathbb
P}}^{(\mbox{\tiny f})}}$ in (\ref{eq:average-work0})) expressed in
terms of the ``dissipative heat''
\begin{equation}\label{eq:dissipative-heat-0-huimi}
    \mathbb{Q}_{\scriptscriptstyle{\Upsilon},\mbox{\scriptsize{d}}}^{(\mbox{\scriptsize f})} = \frac{1}{\beta}\, \left\{\frac{{\mathbb K}_{\tau}}{\widetilde{{\mathbb K}}_{\tau} +
    a_{\scriptscriptstyle{\Upsilon}}\,\tanh(\beta\hbar\omega_0/2)} - \frac{1}{1 + a_{\scriptscriptstyle{\Upsilon}}\,\tanh(\beta\hbar\omega_0/2)}
    + \ln\frac{\widetilde{{\mathbb K}}_{\tau} + a_{\scriptscriptstyle{\Upsilon}}\,\tanh(\beta\hbar\omega_0/2)}{1 +
    a_{\scriptscriptstyle{\Upsilon}}\,\tanh(\beta\hbar\omega_0/2)}\right\}\,.
\end{equation}
Then it follows that
$\mathbb{Q}_{\mbox{\tiny{Q}},\mbox{\scriptsize{d}}}^{(\mbox{\scriptsize
f})} \ngeq 0$ displays the violation of the second law, and
$\mathbb{Q}_{\mbox{\tiny{P}},\mbox{\scriptsize{d}}}^{(\mbox{\scriptsize
f})}$ incorrectly behaves in the low-temperature regime (cf. Fig.
\ref{fig:fig5}).

As a result, we see that although the internal energy and its
difference can be uniquely determined regardless of the phase-space
representations under consideration, it is the Wigner representation
only that propagates in the most classical way among them.
Therefore, it is legitimate to say that the Wigner representation is
the most appropriate choice for the study of the average
thermodynamic work in the quantum regime equipped with the canonical
transition to its classical counterpart $\<W\> = \int_0^{\tau} dt\,
\dot{\lambda}\, \partial_{\lambda} H(z_t;\lambda_t)$ in the simplest
way. In fact, thermodynamics originally appeared from the classical
domain. Therefore, defining quantum work in the Wigner
representation is a consistent step toward a generalization of the
classical work.

%%%%%%%%%%%%%%%%%%%%%%%%%%%%%%%%%%%%%%%%%%%%%%%%%%%%%%%%%%%%%%%%%%%%%%%%%%%%%
\subsection{Comments on our results}\label{subsec:comments}
%%%%%%%%%%%%%%%%%%%%%%%%%%%%%%%%%%%%%%%%%%%%%%%%%%%%%%%%%%%%%%%%%%%%%%%%%%%%%
%
Several additional comments are deserved here. First, it is
instructive to compare our exact analysis with the semiclassical
analysis, carried out in \cite{JAR15}, in the context of the
quantum-classical correspondence principle. In their analysis, the
classical work distribution [cf. Eq.
(\ref{eq:prob-density1-classical-f})], built from the classical
trajectories that connect the initial and final energies, has shown
an excellent approximation to its TPM counterpart in the
semiclassical regime. On the other hand, our work distribution in
the quantum regime, built from the Wigner trajectories that connect
the initial and final action values, renders the average quantum
work distinguished from the internal energy difference; however, in
the classical limit, as discussed above, these trajectories exactly
reduce to the classical ones and this distinguishment between the
average work and internal energy difference goes away. Consequently,
we may say that our quantal-classical analysis consistently
accommodates this semiclassical analysis of the TPM framework.

Second, we add remarks upon the adiabatic process (${\mathbb
K}_{\tau} = 1$): If an external driving acts infinitely slowly,
there are no transitions between different eigenstates (cf.
\cite{BOR25,DIT16} for the quantum adiabatic theorem) such that the
conditional probability $P[m(\tau)|n(0)] = \delta_{nm}$, and thus no
additional quantum fluctuation is produced over the driving.
Therefore, while in the non-adiabatic process the quantum work
${\mathrm w}_{nm}$ of the TPM framework has no independent physical
reality until completion of the final measurement \cite{JAR15}, in
the adiabatic process this work has the physical reality over the
driving indeed; therefore, in this case, the stochastic nature of
the quantum work becomes associated solely with the random nature of
the initial state, like in the classical setup. Then, the quantum
heat $\mathbb{Q}_{\mbox{\scriptsize q}}^{(\mbox{\scriptsize f})}$ in
the Wigner representation also appears solely from the quantum
fluctuation in the initial state.

Third, it is also instructive to emphasize that we resorted to the
action variable ($\mathbb{I}$) in our discussion, not to the EBK
quantization rule given by $\mathbb{I}_n = \hbar\,(n + \alpha/4)$
\cite{BRA97,DIT16}: For a linear oscillator with the Maslov index
$\alpha = 2$, this semiclassical quantization is exactly valid over
the entire quantum regime such that the energy eigenvalue $E_n =
\omega\,\mathbb{I}_n$. On the other hand, our approach simply made
use of the continuous nature of the action variable, which underlies
the continuous nature of the thermodynamic work
$\mathbb{W}_{(\scriptscriptstyle \mathbb{I}_0,\mathbb{I}_{\tau})}$
in the quantum regime.

Finally, we remind that the work distribution in the Wigner
representation is positive valued for a linear oscillator because of
the Gaussian nature of the initial thermal state and its time
evolution [cf. Eqs.
(\ref{eq:proba-forward1})-(\ref{eq:proba-backward1})]. However, this
is not the case any longer for generic systems such as a single
particle confined by a one-dimensional infinite potential. Further,
in our framework of quantum thermodynamics formulated without
resorting to any projective measurements, the measurability of
single-motion values $\Delta e_{(\scriptscriptstyle
\mathbb{I}_0,\mathbb{I}_{\tau})}$ and
$\mathbb{W}_{(\scriptscriptstyle \mathbb{I}_0,\mathbb{I}_{\tau})}$
is inherently abandoned even for the process starting from the
thermal state because of the quasi-probabilistic nature of the
Wigner function; instead, our concern lies in the average values
$\Delta U(\tau)$ and $\<\mathbb{W}\>_{{\scriptscriptstyle {\mathbb
P}}}$ only that reveal the more fine-grained form of the quantum
second law in (\ref{eq:finer-form-0}) with the canonical transition
to its classical counterpart. In fact, the TPM framework is viewed
as the special case only that the measurability of single-run values
$\Delta e_{nm}$ is available. However, it is such a measurement-free
nature of our framework that enables us to be free from a
determination of the energy eigenvalues (required for the TPM
framework) and straightforwardly generalize our findings to the
processes starting from the non-thermal initial states with
coherence in the energy basis. This subject will explicitly be
covered by the following section.

%%%%%%%%%%%%%%%%%%%%%%%%%%%%%%%%%%%%%%%%%%%%%%%%%%%%%%%%%%%%%%%%%%%%%%%%%%%%%
\section{Partially Thermal Initial States for Fluctuation Relations}\label{sec:non-thermal}
%%%%%%%%%%%%%%%%%%%%%%%%%%%%%%%%%%%%%%%%%%%%%%%%%%%%%%%%%%%%%%%%%%%%%%%%%%%%%
%
Now we generalize the Crooks fluctuation theorem in Eq.
(\ref{eq:crooks-quantum1}) by considering the partially thermal
states (as a particular class of non-equilibrium initial states)
such that $\hat{\rho}_0 = (1-\gamma)\,\hat{\rho}_{\beta}(\omega_0) +
\gamma\,\hat{\sigma}(\omega_0)$  for the forward process and
$\hat{\rho}_{\tau} = (1-\gamma)\,\hat{\rho}_{\beta}(\omega_{\tau}) +
\gamma\,\hat{\sigma}(\omega_{\tau})$ for the backward process. Here,
the symbols $\hat{\sigma}$ and $\gamma$ (with $0 \leq \gamma < 1/2$)
denote a non-thermal state and an imperfection in preparing the
thermal state $\hat{\rho}_{\beta}$, respectively. To do so, we first
modify $A^{(\mbox{\scriptsize f})}(0)$ in Eq.
(\ref{eq:phase-space1}), simply by replacing $W_{\beta}$ with
$W_{\rho_0}$, into
\begin{equation}\label{eq:dist2-0}
    A_{\rho_{\scriptscriptstyle 0}}^{(\mbox{\scriptsize f})}(0) = \int dx dp \int dx' dp'\;
    W_{\rho_{\scriptscriptstyle 0}}(x',p';\omega_0)\; T_{\mbox{\tiny W}}^{(\mbox{\scriptsize
    f})}(x,p;\tau|x',p';0)\,,
\end{equation}
being unity. Then it is straightforward to rewrite this as
$A_{\rho_{\scriptscriptstyle 0}}^{(\mbox{\scriptsize f})}(0) =
\int_0^{\infty}\int_0^{\infty} d\mathbb{I}_0\, d\mathbb{I}_{\tau}\;
{\mathbb P}_{\rho_{\scriptscriptstyle 0}}^{(\mbox{\scriptsize
f})}(\mathbb{I}_0,\mathbb{I}_{\tau})$, in which the joint
distribution for the forward process is given by [cf. Eq.
(\ref{eq:proba-forward1})]
%\begin{subequations}
\begin{equation}
    {\mathbb P}_{\rho_{\scriptscriptstyle 0}}^{(\mbox{\scriptsize
    f})}(\mathbb{I}_0,\mathbb{I}_{\tau}) =
    (1-\gamma)\; {\mathbb P}^{(\mbox{\scriptsize
    f})}(\mathbb{I}_0,\mathbb{I}_{\tau}) +
    \gamma\; G_{\sigma}^{(\mbox{\scriptsize f})}(\mathbb{I}_0,\mathbb{I}_{\tau})\,.\label{eq:dist2}
\end{equation}
Here, the second term on the right-hand side will be determined
explicitly for a given state $\hat{\sigma}$. Likewise, we have for
the backward process [cf. Eq. (\ref{eq:proba-backward1})]
\begin{equation}
    {\mathbb P}_{\rho_{\scriptscriptstyle \tau}}^{(\mbox{\scriptsize
    b})}(\mathbb{I}_{\tau},\mathbb{I}_0) =
    (1-\gamma)\; {\mathbb P}^{(\mbox{\scriptsize
    b})}(\mathbb{I}_{\tau},\mathbb{I}_0) +
    \gamma\; G_{\sigma}^{(\mbox{\scriptsize b})}(\mathbb{I}_{\tau},\mathbb{I}_0)\,.\label{eq:dist2b}
\end{equation}
%\end{subequations}
These two distributions can, in general, be negative valued and will
be used for the generalized Crooks theorem in the Wigner
representation. Several particular cases for $\hat{\sigma}$ will be
under consideration below.

%%%%%%%%%%%%%%%%%%%%%%%%%%%%%%%%%%%%%%%%%%%%%%%%%%%%%%%%%%%%%%%%%%%%%%%%%%%%%
%\section{Particular Non-thermal Initial States}\label{sec:non-thermal}
%%%%%%%%%%%%%%%%%%%%%%%%%%%%%%%%%%%%%%%%%%%%%%%%%%%%%%%%%%%%%%%%%%%%%%%%%%%%%
%
%%%%%%%%%%%%%%%%%%%%%%%%%%%%%%%%%%%%%%%%%%%%%%%%%%%%%%%%%%%%%%%%%%%%%%%%%%%%%
\subsection{Mixture of thermal and eigen-energy states}\label{subsec:non-thermal1}
%%%%%%%%%%%%%%%%%%%%%%%%%%%%%%%%%%%%%%%%%%%%%%%%%%%%%%%%%%%%%%%%%%%%%%%%%%%%%
%
Let $\hat{\sigma}_1 = |n\>\<n|$ with
$W_{\sigma_{\scriptscriptstyle{1}}}(x,p)$. Then, we need the Wigner
function of the $n$th energy eigenstate in Eq.
(\ref{eq:wignerfkt-eigen1}); e.g.,
\begin{equation}\label{eq:dist7}
    W_0(\mathbb{I}) = \frac{1}{\pi\hbar}\,e^{-2\,\mathbb{I}/\hbar}\,;\, W_1(\mathbb{I}) =
    \frac{1}{\pi\hbar}\,e^{-2\,\mathbb{I}/\hbar}\,\left(\frac{4\,\mathbb{I}}{\hbar} - 1\right)\,;\, W_2(\mathbb{I}) =
    \frac{1}{\pi\hbar}\,e^{-2\,\mathbb{I}/\hbar}\,\left(\frac{8\,\mathbb{I}^2}{\hbar^2} - \frac{8\,\mathbb{I}}{\hbar} +
    1\right)\,.
\end{equation}
The two quasi-probability distributions $W_1(\mathbb{I})$ and
$W_2(\mathbb{I})$ can be negative valued indeed. By combining Eqs.
(\ref{eq:dist2}) and (\ref{eq:dist2b}) with
$G_{\sigma_{\scriptscriptstyle 1}}^{(\mbox{\scriptsize f,b})} \to
W_n^{(\mbox{\scriptsize f,b})} B^{(\mbox{\scriptsize f})}$ in this
case, it is straightforward to acquire the generalized Crooks
fluctuation theorem
\begin{equation}\label{eq:dist14}
    \frac{{\mathbb P}_{\underline{1}}^{(\mbox{\scriptsize
    f})}(\mathbb{I}_0,\mathbb{I}_{\tau})}{{\mathbb P}_{\underline{1}}^{(\mbox{\scriptsize
    b})}(\mathbb{I}_{\tau},\mathbb{I}_0)} = \frac{(1 -
    \gamma)\, W_{\beta}(\mathbb{I}_{\tau};\omega_{\tau}) + \gamma\, W_n(\mathbb{I}_{\tau})}{(1 -
    \gamma)\, W_{\beta}(\mathbb{I}_0;\omega_0) + \gamma\, W_n(\mathbb{I}_0)}
    = \exp\left\{\beta \left(\mathbb{W}_{\underline{1},(\scriptscriptstyle \mathbb{I}_0,\mathbb{I}_{\tau})} -
    \Delta F_{\beta}^{(\mbox{\scriptsize f})}\right)\right\} \geq 0
\end{equation}
(${\mathbb P}_{\rho_{\scriptscriptstyle 0,1}}^{(\mbox{\scriptsize
f})} \to {\mathbb P}_{\underline{1}}^{(\mbox{\scriptsize f})}$ and
${\mathbb P}_{\rho_{\scriptscriptstyle \tau,1}}^{(\mbox{\scriptsize
b})} \to {\mathbb P}_{\underline{1}}^{(\mbox{\scriptsize b})}$ in
notation); here, the generalized work is identified as
\begin{equation}\label{eq:phase-space7-sig}
    \mathbb{W}_{\underline{1},(\scriptscriptstyle \mathbb{I}_0,\mathbb{I}_{\tau})}
    = \mathbb{W}_{(\scriptscriptstyle \mathbb{I}_0,\mathbb{I}_{\tau})} - \frac{1}{\beta}\,
    \ln\left[\frac{1 + \{\gamma/(1 -
    \gamma)\}\,
    W_n(\mathbb{I}_{\tau})/W_{\beta}(\mathbb{I}_{\tau};\omega_{\tau})}{1 + \{\gamma/(1 -
    \gamma)\}\,
    W_n(\mathbb{I}_0)/W_{\beta}(\mathbb{I}_0;\omega_0)}\right]
\end{equation}
for a given value of $\gamma$ with $\gamma/(1 - \gamma) < 1$ [cf.
Eq. (\ref{eq:phase-space7})]. As such, the second term is not linear
in $\mathbb{I}_0$ and $\mathbb{I}_{\tau}$ any longer. We also find
that the initial state $W_{\rho_{\scriptscriptstyle
0,1}}(x,p;\omega_0)$ and so the distribution ${\mathbb
P}_{\underline{1}}^{(\mbox{\scriptsize
f})}(\mathbb{I}_0,\mathbb{I}_{\tau})$ can be negative valued indeed
if $n \ne 0$ and $\gamma_{\mbox{\scriptsize
th},\scriptscriptstyle{\underline{1}}} < \gamma < 1/2$, where the
threshold value $\gamma_{\mbox{\scriptsize
th},\scriptscriptstyle{\underline{1}}}$ should be determined for the
respective initial state (cf. Fig. \ref{fig:fig6}).

Then, Eq. (\ref{eq:dist14}) will yield the Jarzynski equality
\begin{equation}\label{eq:dist15-0-0}
    \left\<e^{-\beta \mathbb{W}}\right\>_{{\scriptscriptstyle {\mathbb P}}^{(\mbox{\tiny f})}_{{\scriptscriptstyle\underline{1}}}}
    = \int d\mathbb{W}\; e^{-\beta\,\mathbb{W}}\;
    {\mathbb P}_{\underline{1}}^{(\mbox{\scriptsize f})}(\mathbb{W}) =
    e^{-\beta \Delta F_{\beta}^{(\mbox{\tiny f})}}
\end{equation}
where the work distribution is
\begin{equation}\label{eq:work-dist-new_1-sig}
    {\mathbb P}_{\underline{1}}^{(\mbox{\scriptsize f})}(\mathbb{W}) = \int_0^{\infty}\int_0^{\infty}
    d\mathbb{I}_0\, d\mathbb{I}_{\tau}\; \delta(\mathbb{W} - \mathbb{W}_{\underline{1},(\scriptscriptstyle \mathbb{I}_0,\mathbb{I}_{\tau})})\; {\mathbb P}_{\underline{1}}^{(\mbox{\scriptsize
    f})}(\mathbb{I}_0,\mathbb{I}_{\tau})\,.
\end{equation}
Eq. (\ref{eq:dist15-0-0}) will finally give rise to the generalized
second-law inequality
\begin{equation}\label{eq:dist15-0}
    \Delta F_{\beta}^{(\mbox{\scriptsize f})} \leq \left\<\mathbb{W}\right\>_{{\scriptscriptstyle {\mathbb P}}_{\underline{1}}^{(\mbox{\tiny f})}} \leq \Delta
    U_{\underline{1}}(\tau)
\end{equation}
[cf. (\ref{eq:finer-form-0}) for $r = 0$]. Here, the internal energy
difference
\begin{equation}\label{eq:internal-energy_diff10}
    \Delta U_{\underline{1}}(\tau) = \omega_{\tau}\,\<\mathbb{I}_{\tau}\>_{{\scriptscriptstyle {\mathbb P}}_{\underline{1}}^{(\mbox{\tiny f})}} -
    \omega_0\,\<\mathbb{I}_0\>_{{\scriptscriptstyle {\mathbb P}}_{\underline{1}}^{(\mbox{\tiny f})}} =
    (\hbar/2)\, ({\omega_{\tau}\,\mathbb K}_{\tau} - \omega_0)\, \{(1-r)\, \mbox{coth}(\beta \hbar \omega_0/2)\, +\, r\, (2n +
    1)\}
\end{equation}
is still expressed in terms of both first moments [cf.
(\ref{eq:phase-space-12})]. On the other hand, the average work
$\left\<\mathbb{W}\right\>_{{\scriptscriptstyle {\mathbb
P}}_{\underline{1}}^{(\mbox{\tiny f})}}$ will be expressed in terms
of the higher-order moments in addition to the first moments; cf.
Eqs.
(\ref{eq:higher-order-moments-1})-(\ref{eq:higher-order-moments-3})
and (\ref{eq:3rd-moments-1})-(\ref{eq:3rd-moments-2}) are useful
also for $r \ne 0$ in this case. The second inequality in
(\ref{eq:dist15-0}) can be verified, as in
(\ref{eq:second-law-first-part1}) for $r=0$. These behaviors of
$\left\<\mathbb{W}\right\>_{{\scriptscriptstyle {\mathbb
P}}_{\underline{1}}^{(\mbox{\tiny f})}}$ are demonstrated in Fig.
\ref{fig:fig7}.

%%%%%%%%%%%%%%%%%%%%%%%%%%%%%%%%%%%%%%%%%%%%%%%%%%%%%%%%%%%%%%%%%%%%%%%%%%%%%
\subsection{Mixture of thermal and energy-superposed states: Case 1}\label{subsec:non-thermal2}
%%%%%%%%%%%%%%%%%%%%%%%%%%%%%%%%%%%%%%%%%%%%%%%%%%%%%%%%%%%%%%%%%%%%%%%%%%%%%
%
Let $\hat{\sigma}_2 = |n\> + |n+1\>)(\<n| + \<n+1|)/2$, which
possesses the energy coherence. Then, we also need the Moyal
functions \cite{BUZ95}
\begin{equation}\label{eq:moyal1}
    W_{|m\>\<n|}(x,p) = \frac{(-1)^n}{\pi\hbar}
    \left(\frac{n!}{m!}\right)^{1/2} \{2\,\eta^{\ast}(x,p)\}^{m-n}\; e^{-2\,|\eta(x,p)|^2}\;
    L_n^{(|m-n|)}(4\,|\eta(x,p)|^2)
\end{equation}
for $n \leq m$, where $L_n^{(k)}(\cdots)$ denotes the associated
Laguerre polynomial, and $W_{|n\>\<m|}(x,p) =
\{W_{|m\>\<n|}(x,p)\}^{\ast}$; e.g.,
\begin{subequations}
\begin{eqnarray}
    \hspace*{-.7cm}W_{|1\>\<0|}(x,p) &=& \frac{\sqrt{2}}{\pi \hbar}\, \left(\kappa x -
    \frac{i p}{\hbar\kappa}\right)\, \exp\left[-\left\{(\kappa
    x)^2 + \left(\frac{p}{\hbar\kappa}\right)^2\right\}\right]\label{eq:off-diagonal-wig1}\\
    \hspace*{-.7cm}W_{|2\>\<1|}(x,p) &=& \frac{2}{\pi \hbar}\, \left(\kappa x -
    \frac{i p}{\hbar\kappa}\right)\, \left\{(\kappa
    x)^2 + \left(\frac{p}{\hbar\kappa}\right)^2 - 1\right\}\, \exp\left[-\left\{(\kappa
    x)^2 + \left(\frac{p}{\hbar\kappa}\right)^2\right\}\right]\label{eq:dist17-2}\\
    \hspace*{-.7cm}W_{|2\>\<0|}(x,p) &=& \frac{\sqrt{2}}{\pi \hbar}\, \left\{(\kappa x)^2 -
    \left(\frac{p}{\hbar\kappa}\right)^2 - \frac{2i\,x p}{\hbar}\right\}\, \exp\left[-\left\{(\kappa
    x)^2 + \left(\frac{p}{\hbar\kappa}\right)^2\right\}\right]\,.\label{eq:dist18}
\end{eqnarray}
\end{subequations}
We see that the angle coordinate $\phi$ will also be needed for the
Moyal functions $W_{|m\>\<n|}(\phi,\mathbb{I})$. Employing Eqs.
(\ref{eq:wignerfkt-eigen1}) and (\ref{eq:moyal1}), we can obtain the
Wigner function
\begin{equation}\label{eq:sigma_2_state0}
    W_{\sigma_2}(\phi,\mathbb{I};n) = \frac{(-1)^n}{2\pi\hbar}\, e^{-2\,\mathbb{I}/\hbar}\, \left\{L_n(4\,\mathbb{I}/\hbar) -
    L_{n+1}(4\,\mathbb{I}/\hbar) + 4
    \sqrt{\mathbb{I}/\hbar\,(n+1)}\, (\sin\phi)\,
    L_n^{(1)}(4\,\mathbb{I}/\hbar)\right\}\,.
\end{equation}

Then, we can find from Eq. (\ref{eq:dist2}) that the joint
distribution for the forward process is
\begin{eqnarray}\label{eq:dist11f-2}
    {\mathbb P}_{\underline{2}}^{(\mbox{\scriptsize
    f})}(\mathbb{I}_0,\mathbb{I}_{\tau}) &=&
    B^{(\mbox{\scriptsize f})}(\mathbb{I}_{\tau}|\mathbb{I}_0)\, \left\{(1-\gamma)\, W_{\beta}(\mathbb{I}_0;\omega_0) + \gamma\,
    \frac{W_n(\mathbb{I}_0) + W_{n+1}(\mathbb{I}_0)}{2}\right\} +\n\\
    && \gamma\, \frac{2\,(-1)^n}{\pi\hbar}\, C^{(\mbox{\scriptsize
    f})}(\mathbb{I}_{\tau}|\mathbb{I}_0)\,
    \left\{\frac{\mathbb{I}_0}{\hbar\,(n+1)}\right\}^{1/2}\,
    e^{-2\,\mathbb{I}_0/\hbar}\;
    L_n^{(1)}(4\,\mathbb{I}_0/\hbar)\,,
\end{eqnarray}
where the second conditional distribution is given by
\begin{equation}
    C^{(\mbox{\scriptsize f})}(\mathbb{I}_{\tau}|\mathbb{I}_0) = \mbox{Re} \int_0^{2\pi}\int_0^{2\pi}
    d\phi_{\tau}\, d\phi_0\; (\sin\phi_0)\; \tilde{T}_{\mbox{\tiny{W}}}^{(\mbox{\scriptsize
    f})}(\phi_{\tau},\mathbb{I}_{\tau};\tau|\phi_0,\mathbb{I}_0;0)\,.\label{eq:phase-space3-3-C0}
\end{equation}
In fact, we can show that $C^{(\mbox{\scriptsize
f})}(\mathbb{I}_{\tau}|\mathbb{I}_0) \equiv 0$ (cf. Appendix
\ref{sec:appendix2}). Consequently, we see that the off-diagonal
terms of the initial state $\hat{\rho}_{0,2}$ do not contribute to
${\mathbb P}_{\underline{2}}^{(\mbox{\scriptsize
f})}(\mathbb{I}_0,\mathbb{I}_{\tau})$, meaning that the diagonal
form $\utilde{\hat{{\sigma}}}_2 = (|n\>\<n| + |n+1\>\<n+1|)/2$, in
place of $\hat{\sigma}_2$, will give the same result for ${\mathbb
P}_{\underline{2}}^{(\mbox{\scriptsize
f})}(\mathbb{I}_0,\mathbb{I}_{\tau})$. Similarly, we can acquire the
joint distribution for the backward process
\begin{equation}\label{eq:dist113-2}
    {\mathbb P}_{\underline{2}}^{(\mbox{\scriptsize
    b})}(\mathbb{I}_{\tau},\mathbb{I}_0) =
    B^{(\mbox{\scriptsize f})}(\mathbb{I}_{\tau}|\mathbb{I}_0)\, \left\{(1-\gamma)\; W_{\beta}(\mathbb{I}_{\tau};\omega_{\tau}) +
    \gamma\; \frac{W_n(\mathbb{I}_{\tau}) + W_{n+1}(\mathbb{I}_{\tau})}{2}\right\}\,.
\end{equation}
Combining Eqs. (\ref{eq:dist11f-2}) and (\ref{eq:dist113-2}), it is
straightforward to acquire the Crooks fluctuation theorem in the
form of Eq. (\ref{eq:dist14}) and the Jarzynski equality in the form
of (\ref{eq:dist15-0-0}) as well as the second-law inequality in the
form of (\ref{eq:dist15-0}), where the pertinent work
$\mathbb{W}_{\underline{2},(\scriptscriptstyle
\mathbb{I}_0,\mathbb{I}_{\tau})}$ is accordingly given by Eq.
(\ref{eq:phase-space7-sig}) but with $W_n(\mathbb{I}_t) \to
\{W_n(\mathbb{I}_t) + W_{n+1}(\mathbb{I}_t)\}/2$, and its
distribution is then
\begin{equation}\label{eq:work-dist-new_1-sig-11}
    {\mathbb P}_{\underline{2}}^{(\mbox{\scriptsize f})}(\mathbb{W}) = \int_0^{\infty}\int_0^{\infty}
    d\mathbb{I}_0\, d\mathbb{I}_{\tau}\; \delta(\mathbb{W} -
    \mathbb{W}_{\underline{2},(\scriptscriptstyle \mathbb{I}_0,\mathbb{I}_{\tau})})\; {\mathbb P}_{\underline{2}}^{(\mbox{\scriptsize
    f})}(\mathbb{I}_0,\mathbb{I}_{\tau})\,.
\end{equation}
Also, the internal energy difference $\Delta U_{\underline{1}}(\tau)
\to \Delta U_{\underline{2}}(\tau)$ such that
\begin{equation}\label{eq:internal-energy_diff11}
    \Delta U_{\underline{2}}(\tau) = \omega_{\tau}\,\<\mathbb{I}_{\tau}\>_{{\scriptscriptstyle {\mathbb P}}_{\underline{2}}^{(\mbox{\tiny f})}} -
    \omega_0\,\<\mathbb{I}_0\>_{{\scriptscriptstyle {\mathbb P}}_{\underline{2}}^{(\mbox{\tiny f})}} =
    (\hbar/2)\, ({\omega_{\tau}\,\mathbb K}_{\tau} - \omega_0)\, \{(1-r)\, \mbox{coth}(\beta \hbar \omega_0/2)\, +\, 2 r\, (n +
    1)\}\,.
\end{equation}

%%%%%%%%%%%%%%%%%%%%%%%%%%%%%%%%%%%%%%%%%%%%%%%%%%%%%%%%%%%%%%%%%%%%%%%%%%%%%
\subsection{Mixture of thermal and energy-superposed states: Case 2}\label{subsec:non-thermal3}
%%%%%%%%%%%%%%%%%%%%%%%%%%%%%%%%%%%%%%%%%%%%%%%%%%%%%%%%%%%%%%%%%%%%%%%%%%%%%
%
Let $\hat{\sigma}_3 = (|n\> + |n+1\> + |n+2\>)(\<n| + \<n+1| +
\<n+2|)/3$. Then, we have
\begin{equation}\label{eq:wig_sig313}
   W_{\sigma_3}(\phi,\mathbb{I}) =  \frac{2}{3}\, \left\{W_{\sigma_2}(\phi,\mathbb{I};n) + W_{\sigma_2}(\phi,\mathbb{I};n+1)\right\}
   + \frac{1}{3}\, \left\{-W_{n+1}(\phi,\mathbb{I}) + F_{n,n+2}(\phi,\mathbb{I})\right\}
\end{equation}
[cf. Eq. (\ref{eq:sigma_2_state0})], where the last term on the
right-hand side is
\begin{eqnarray}\label{eq:sigma_2_state}
    F_{n,n+2}(\phi,\mathbb{I}) &=& W_{|n+2\>\<n|}(\phi,\mathbb{I}) + W_{|n\>\<n+2|}(\phi,\mathbb{I})\\
    &=& \frac{8\,(-1)^n}{\pi\hbar}\, \sqrt{\{(n+1) (n+2)\}^{-1}}\; e^{-2\,\mathbb{I}/\hbar}\;
    (\mathbb{I}/\hbar)\;
    \{2\,(\sin\phi)^2 - 1\}\; L_n^{(2)}(4\,\mathbb{I}/\hbar)\n\,.
\end{eqnarray}
We can obtain from Eq. (\ref{eq:dist2}) the joint distribution for
the forward process
\begin{equation}\label{eq:dist11f-2-1}
    {\mathbb P}_{\underline{3}}^{(\mbox{\scriptsize
    f})}(\mathbb{I}_0,\mathbb{I}_{\tau}) =
    (1-\gamma)\; B^{(\mbox{\scriptsize f})}(\mathbb{I}_{\tau}|\mathbb{I}_0)\; W_{\beta}(\mathbb{I}_0;\omega_0) +
    \gamma\; G_{\underline{3}}^{(\mbox{\scriptsize
    f})}(\mathbb{I}_0,\mathbb{I}_{\tau})\,,
\end{equation}
where \begin{eqnarray}\label{eq:dist11f-2-1-1}
    &&G_{\underline{3}}^{(\mbox{\scriptsize
    f})}(\mathbb{I}_0,\mathbb{I}_{\tau}) =
    G_{\utilde{3}}^{(\mbox{\scriptsize
    f})}(\mathbb{I}_0,\mathbb{I}_{\tau}) +\\
    &&\frac{8\,(-1)^n}{3\pi\hbar}\;
    \sqrt{\{(n+1) (n+2)\}^{-1}}\; \{2\,D^{(\mbox{\scriptsize
    f})}(\mathbb{I}_{\tau}|\mathbb{I}_0) - B^{(\mbox{\scriptsize
    f})}(\mathbb{I}_{\tau}|\mathbb{I}_0)\}\; (\mathbb{I}_0/\hbar)\;
    e^{-2\,\mathbb{I}_0/\hbar}\;
    L_n^{(2)}(4\,\mathbb{I}_0/\hbar)\,.\n
\end{eqnarray}
Here, the diagonal-term contribution and the third conditional
distribution are
\begin{subequations}
\begin{eqnarray}
    G_{\utilde{3}}^{(\mbox{\scriptsize
    f})}(\mathbb{I}_0,\mathbb{I}_{\tau}) &=& B^{(\mbox{\scriptsize f})}(\mathbb{I}_{\tau}|\mathbb{I}_0)\; \frac{W_n(\mathbb{I}_0) + W_{n+1}(\mathbb{I}_0) + W_{n+2}(\mathbb{I}_0)}{3}\\
    D^{(\mbox{\scriptsize f})}(\mathbb{I}_{\tau}|\mathbb{I}_0) &=& \mbox{Re} \int_0^{2\pi}\int_0^{2\pi}
    d\phi_{\tau}\, d\phi_0\; (\sin\phi_0)^2\; \tilde{T}_{\mbox{\tiny{W}}}^{(\mbox{\scriptsize
    f})}(\phi_{\tau},\mathbb{I}_{\tau};\tau|\phi_0,\mathbb{I}_0;0)\,,\label{eq:phase-space3-3-C}
\end{eqnarray}
\end{subequations}
respectively (cf. Appendix \ref{sec:appendix2}). We see that the
joint distribution ${\mathbb P}_{\underline{3}}^{(\mbox{\scriptsize
f})}$ differs from its counterpart ${\mathbb
P}_{\utilde{3}}^{(\mbox{\scriptsize f})}$ given by Eq.
(\ref{eq:dist11f-2-1}) but with
$G_{\underline{3}}^{(\mbox{\scriptsize f})} \to
G_{\utilde{3}}^{(\mbox{\scriptsize f})}$ (without coherence),
obtained from the diagonal form $\utilde{\hat{{\sigma}}}_3 =
(|n\>\<n| + |n+1\>\<n+1| + |n+2\>\<n+2|)/3$ in place of
$\hat{\sigma}_3$. Likewise, the joint distribution for the backward
process is given by
\begin{equation}\label{eq:dist11f-2-1b}
    {\mathbb P}_{\underline{3}}^{(\mbox{\scriptsize
    b})}(\mathbb{I}_{\tau},\mathbb{I}_0) =
    (1-\gamma)\; B^{(\mbox{\scriptsize f})}(\mathbb{I}_{\tau}|\mathbb{I}_0)\; W_{\beta}(\mathbb{I}_{\tau};\omega_{\tau}) +
    \gamma\; G_{\underline{3}}^{(\mbox{\scriptsize
    b})}(\mathbb{I}_{\tau},\mathbb{I}_0)\,,
\end{equation}
in which $G_{\underline{3}}^{(\mbox{\scriptsize
b})}(\mathbb{I}_{\tau},\mathbb{I}_0)$ is given by Eq.
(\ref{eq:dist11f-2-1-1}) but with $\mathbb{I}_0 \to
\mathbb{I}_{\tau}$ while $B^{(\mbox{\scriptsize
f})}(\mathbb{I}_{\tau}|\mathbb{I}_0)$ remains unaffected, and
$D^{(\mbox{\scriptsize b})}(\mathbb{I}_0|\mathbb{I}_{\tau})$ is
given by Eq. (\ref{eq:phase-space3-3-C}) but with $\sin\phi_0 \to
\sin\phi_{\tau}$.

Then it is easy to acquire the Crooks fluctuation theorem, the
Jarzynski equality and the second-law inequality in the form of Eqs.
(\ref{eq:dist14}), (\ref{eq:dist15-0-0}) and (\ref{eq:dist15-0}),
respectively. Here, the pertinent work
$\mathbb{W}_{\underline{3},(\scriptscriptstyle
\mathbb{I}_0,\mathbb{I}_{\tau})}$ is explicitly given by Eq.
(\ref{eq:phase-space7-sig}) but with $W_n(\mathbb{I}_0) \to
G_{\underline{3}}^{(\mbox{\scriptsize
f})}(\mathbb{I}_0,\mathbb{I}_{\tau})/B^{(\mbox{\scriptsize
f})}(\mathbb{I}_{\tau}|\mathbb{I}_0)$ and $W_n(\mathbb{I}_{\tau})
\to G_{\underline{3}}^{(\mbox{\scriptsize
b})}(\mathbb{I}_{\tau},\mathbb{I}_0)/B^{(\mbox{\scriptsize
f})}(\mathbb{I}_{\tau}|\mathbb{I}_0)$, as well as its distribution
is
\begin{equation}\label{eq:work-dist-new_1-sig-11-1}
    {\mathbb P}_{\underline{3}}^{(\mbox{\scriptsize f})}(\mathbb{W}) = \int_0^{\infty}\int_0^{\infty}
    d\mathbb{I}_0\, d\mathbb{I}_{\tau}\; \delta(\mathbb{W} -
    \mathbb{W}_{\underline{3},(\scriptscriptstyle \mathbb{I}_0,\mathbb{I}_{\tau})})\; {\mathbb P}_{\underline{3}}^{(\mbox{\scriptsize
    f})}(\mathbb{I}_0,\mathbb{I}_{\tau})\,.
\end{equation}
Also, the internal energy difference $\Delta U_{\underline{1}}(\tau)
\to \Delta U_{\underline{3}}(\tau)$. In Fig. \ref{fig:fig8}, the
behaviors of $\left\<\mathbb{W}\right\>_{{\scriptscriptstyle
{\mathbb P}}_{\underline{3}}^{(\mbox{\tiny f})}}$ are explicitly
compared with those of its counterpart
$\left\<\mathbb{W}\right\>_{{\scriptscriptstyle {\mathbb
P}}_{\utilde{3}}^{(\mbox{\tiny f})}}$ (obtained from ${\mathbb
P}_{\utilde{3}}^{(\mbox{\scriptsize f})}$); as shown, the quantities
$\left\<\mathbb{W}\right\>_{{\scriptscriptstyle {\mathbb
P}}_{\underline{3}}^{(\mbox{\tiny f})}}$ and $\Delta
U_{\underline{3}}(\tau)$ with coherence are greater than their
counterparts $\left\<\mathbb{W}\right\>_{{\scriptscriptstyle
{\mathbb P}}_{\utilde{3}}^{(\mbox{\tiny f})}}$ and $\Delta
U_{\utilde{3}}(\tau)$ without coherence, respectively, where for $n
= 0$
\begin{subequations}
\begin{eqnarray}\label{eq:internal-energy_diff12}
    \Delta U_{\underline{3}}(\tau) &=& \omega_{\tau}\,\<\mathbb{I}_{\tau}\>_{{\scriptscriptstyle {\mathbb P}}_{\underline{3}}^{(\mbox{\tiny f})}} -
    \omega_0\,\<\mathbb{I}_0\>_{{\scriptscriptstyle {\mathbb P}}_{\underline{3}}^{(\mbox{\tiny f})}} =
    \omega_{\tau}\,\left[(1-r)\, \<\mathbb{I}_{\tau}\>_{{\scriptscriptstyle {\mathbb P}}^{(\mbox{\tiny f})}} + \frac{\gamma}{3}\,
    \left\{\frac{8}{\hbar^2}\, \<\mathbb{I}_{\tau}\,(\mathbb{I}_0)^2\>_{{\scriptscriptstyle {\mathbb P}}^{(\mbox{\tiny f})}}\, +\right.\right.\n\\
    && \left.\left.\frac{4}{\hbar}\, (2^{1/2} - 1)\,
    \<\mathbb{I}_{\tau}\,\mathbb{I}_0\>_{{\scriptscriptstyle {\mathbb P}}^{(\mbox{\tiny f})}} +
    \<\mathbb{I}_{\tau}\>_{{\scriptscriptstyle {\mathbb P}}^{(\mbox{\tiny f})}}\right\}_{\beta \to \infty}\right] - \omega_0\; [(\omega_{\tau} \to \omega_0)]\n\\
    &=&
    \frac{\hbar}{2}\, ({\omega_{\tau}\,\mathbb K}_{\tau} - \omega_0)\,
    \left\{(1-r)\, \mbox{coth}(\beta \hbar \omega_0/2)\, +\, \frac{r}{3}\, (9 + 4\,\sqrt{2})\right\}
\end{eqnarray}
[cf. (\ref{eq:3rd-moments-1})-(\ref{eq:3rd-moments-2})], and
\begin{equation}\label{eq:internal-energy_diff12-0}
    \Delta U_{\utilde{3}}(\tau) = \omega_{\tau}\,\<\mathbb{I}_{\tau}\>_{{\scriptscriptstyle {\mathbb P}}_{\utilde{3}}^{(\mbox{\tiny f})}} -
    \omega_0\,\<\mathbb{I}_0\>_{{\scriptscriptstyle {\mathbb P}}_{\utilde{3}}^{(\mbox{\tiny f})}} =
    (\hbar/2)\, ({\omega_{\tau}\,\mathbb K}_{\tau} - \omega_0)\, \{(1-r)\, \mbox{coth}(\beta \hbar \omega_0/2)\, +\, r\, (2n + 3)\}\,.
\end{equation}
\end{subequations}

%%%%%%%%%%%%%%%%%%%%%%%%%%%%%%%%%%%%%%%%%%%%%%%%%%%%%%%%%%%%%%%%%%%%%%%%%%%%%
\subsection{Comments on our results}\label{subsec:comments2}
%%%%%%%%%%%%%%%%%%%%%%%%%%%%%%%%%%%%%%%%%%%%%%%%%%%%%%%%%%%%%%%%%%%%%%%%%%%%%
%
Now we give the interpretation of our findings in the present
section. Differing from the thermal initial state (with $\gamma =
0$), the partially thermal initial state (with $\gamma > 0$) results
in the fact that the internal energy difference is still given by
the first moments, but the average work, obtained from the
generalized work in non-linear form, necessarily contains the
higher-order moment contributions. Remarkably enough, such a
generalized work can also be linked to the fully thermodynamic
quantity $\Delta F_{\beta}$ operationally through our generalized
Crooks theorem (and the resulting Jarzynski equality), finally
giving rise to the second-law inequality associated with the average
work. If the parameter $\gamma$ continues to increase such that it
becomes greater than its threshold value $\gamma_{\mbox{\scriptsize
th}}$ (like in Fig. 6), then a dominance of the quantum fluctuation
over the thermal fluctuation will be found for the initial state in
non-Gaussian form (e.g., $\hat{\sigma}_1 = |n\>\<n|$ with $n \ne 0$)
so that the initial Wigner function and the work distribution can be
negative valued. In the classical scenario, on the other hand, this
exact link between the generalized work and the free energy
difference is well-defined (i.e., with no negativity of the work
probability distribution) also for the single-motion values through
our generalized Crooks theorem in the classical limit. In fact, a
perfect preparation of the thermal state (with $r = 0$) could be a
formidable task in reality.

It is also instructive to point out that our result obtained from
the phase-space framework, free from the projective measurement, is
consistent with the result obtained from the histories framework (as
a different generalization of the TPM framework) in \cite{MIL17},
which employed the time-reversal symmetrized work distributions for
non-thermal initial states, concluding that thermodynamic work in
the quantum regime cannot be determined by the projective
measurements.

%%%%%%%%%%%%%%%%%%%%%%%%%%%%%%%%%%%%%%%%%%%%%%%%%%%%%%%%%%%%%%%%%%%%%%%%%%%%%
\section{Conclusion}\label{sec:conclusion}
%%%%%%%%%%%%%%%%%%%%%%%%%%%%%%%%%%%%%%%%%%%%%%%%%%%%%%%%%%%%%%%%%%%%%%%%%%%%%
%
We studied the quantum fluctuation relations in the Wigner
representation. To make our analysis as exact as possible, we
restricted our discussion here to a driven quantum linear
oscillator. Then we obtained the single-motion work (in closed form)
in the quantum regime and its distribution expressed in terms of the
action coordinates only, without resorting to any projective
measurements, such that the quantum and classical setups can be
analyzed on the single footing. This enabled us to derive the
quantum Crooks fluctuation theorem, the quantum Jarzynski equality
and the second-law inequality in more fine-grained form than their
counterparts in the two-point projective measurement (TPM)
framework, in that the resulting average work $\<\mathbb{W}\>$ in
our framework notably differs from the internal energy difference
$\Delta U(\tau)$ between the initial and final states, thus
rendering the quantum heat $\mathbb{Q}_{\mbox{\scriptsize q}} =
\Delta U(\tau) - \<\mathbb{W}\> \geq 0$ introduced. Such a
discrepancy between $\Delta U(\tau)$ and $\<\mathbb{W}\>$ was shown
to disappear gradually in the semiclassical regime. This result
contrasts with $\<{\mathbb W}\> \equiv \Delta U(\tau$) in the
standard TPM framework. We also provided a justification for the
choice of the Wigner representation for our analysis rather than any
other phase-space representations. We showed that it is the Wigner
representation that behaves in the most classical-like way and is
most appropriate for the quantum work with the canonical transition
to its classical counterpart, in that thermodynamics originally took
its shape from the classical scenario.

Our findings were straightforwardly generalized to the processes
starting from a particular class of non-thermal initial states
including the states with quantum coherence in the eigen-energy
basis. Here, we introduced the generalized quantum work with its
non-linear nature and the work distribution with its negativity. As
a matter of fact, the unavoidable negativity of the work
distribution has been well-known also in the extended TPM framework
where the initial state is non-diagonal in the energy basis. In this
paper, on the other hand, we used such a negativity resulting from
the quasi-probabilistic nature of the Wigner function as our stating
point for a new framework free from the projective measurements,
which led to achieving the aforementioned fine-grained results in
the quantum thermodynamics. As a result, it is legitimate to claim
that our results, covering the genuine quantum to (semi)classical
regimes, can provide a more sophisticated discussion of the second
law of thermodynamics associated with the average work within an
isolated quantum system. Finally, as long as the angle-action
coordinates are well-defined, our methodology will continue to apply
to different quantum systems including the generic one-dimensional
systems.

%%%%%%%%%%%%%%%%%%%%%%%%%%%%%%%%%%%%%%%%%%%%%%%%%%%%%%%%%%%%%%%%%%%
\section*{Acknowledgments}
%%%%%%%%%%%%%%%%%%%%%%%%%%%%%%%%%%%%%%%%%%%%%%%%%%%%%%%%%%%%%%%%%%%
The author gratefully acknowledges the financial support provided by
the US Army Research Office (Grant No. W911NF-15-1-0145).

%%%%%%%%%%%%%%%%%%%%%%%%%%%%%%%%%%%%%%%%%%%%%%%%%%%%%%%%%%%%%%%%%%%%%%%%%%%%%
\appendix\section{Evaluations of Eqs. (\ref{eq:phase-space10}) and (\ref{eq:higher-order-moments-1})-(\ref{eq:higher-order-moments-3})}\label{sec:appendix1}
%%%%%%%%%%%%%%%%%%%%%%%%%%%%%%%%%%%%%%%%%%%%%%%%%%%%%%%%%%%%%%%%%%%%%%%%%%%%%
%
We begin by showing the non-negativity of the conditional
distribution $B^{(\mbox{\scriptsize
f})}(\mathbb{I}_{\tau}|\mathbb{I}_0)$ for all
$(\mathbb{I}_0,\mathbb{I}_{\tau})$ in (\ref{eq:proba-forward1}): We
first substitute Eq. (\ref{eq:propagator-osc1}) into Eq.
(\ref{eq:wigner-propagator2}) with $x = \{2 \mathbb{I}/(m
\omega)\}^{1/2}\,\sin\phi$ and $p = (2 m \omega
\mathbb{I})^{1/2}\,\cos\phi$, followed by executing the integrals
over $\xi_1$ and $\xi_2$. This will transform Eq.
(\ref{eq:phase-space3-3}) into
\begin{equation}\label{eq:appendix_b2-1}
    B^{(\mbox{\scriptsize f})}(\mathbb{I}_{\tau}|\mathbb{I}_0) = \frac{m}{|X_{\tau}|} \int_0^{2\pi}\int_0^{2\pi}
    d\phi_0\, d\phi_{\tau}\;
    \delta\{g_1(\phi_0,\mathbb{I}_0,\phi_{\tau},\mathbb{I}_{\tau})\}\;
    \delta\{g_2(\phi_0,\mathbb{I}_0,\phi_{\tau},\mathbb{I}_{\tau})\}\,,
\end{equation}
where
\begin{equation}\label{eq:appendix-g}
    g_1 = \left(\frac{2 m
    \mathbb{I}_0}{\omega_0}\right)^{1/2}\, \frac{1}{X_{\tau}}\, \sin(\phi_0) + (2 m \omega_{\tau}
    \mathbb{I}_{\tau})^{1/2}\, \cos(\phi_{\tau}) - \left(\frac{2 m
    \mathbb{I}_{\tau}}{\omega_{\tau}}\right)^{1/2}\, \frac{\dot{X}_{\tau}}{X_{\tau}}\,
    \sin(\phi_{\tau})\,,
\end{equation}
and $g_2$ is given by Eq. (\ref{eq:appendix-g}) with ($\phi_0
\leftrightarrow \phi_{\tau}$), ($\mathbb{I}_0 \leftrightarrow
\mathbb{I}_{\tau}$), ($\omega_0 \leftrightarrow \omega_{\tau}$) and
($\dot{X}_{\tau} \to Y_{\tau}$). With the help of the identity
$\delta\{g(y)\} = \sum_j \delta(y - y_j)/|g'(y_j)|$ with $g(y_j) =
0$, we can finally observe that $B^{(\mbox{\scriptsize
f})}(\mathbb{I}_{\tau}|\mathbb{I}_0) \geq 0$ [cf. Eqs
(\ref{eq:wigner-propagator1-1})-(\ref{eq:wigner-propagator1-2})].

Now we explicitly evaluate the first moments
$\<\mathbb{I}_0\>_{{\scriptscriptstyle {\mathbb P}}^{(\mbox{\tiny
f})}}$ and $\<\mathbb{I}_{\tau}\>_{{\scriptscriptstyle {\mathbb
P}}^{(\mbox{\tiny f})}}$ in Eq. (\ref{eq:phase-space10}): For later
purposes, we begin by verifying that $A^{(\mbox{\scriptsize f})}(0)
= \<(\mathbb{I}_{\tau})^0\>_{{\scriptscriptstyle {\mathbb
P}}^{(\mbox{\tiny f})}} = 1$ in (\ref{eq:phase-space3}). With the
help of the relation $(2\pi)\, J_0(\sqrt{A^2 + B^2}) = \int_0^{2\pi}
d\phi\; e^{i\,\{A\,(\sin\phi) + B\,(\cos\phi)\}}$ obtained from the
identity $J_0(y) = (1/\pi) \int_0^{\pi} d\phi\, \cos(y\,\sin\phi)$
for the Bessel function $J_0(y)$ \cite{ABR65}, we can transform Eq.
(\ref{eq:phase-space3-3}) into
\begin{equation}\label{eq:appendix_b1}
    B^{(\mbox{\scriptsize f})}(\mathbb{I}_{\tau}|\mathbb{I}_0) = \frac{m}{\hbar^2\,|X_{\tau}|}
    \int d\xi_1\,d\xi_2\; J_0(b_{\tau} \sqrt{\mathbb{I}_{\tau}})\; J_0(b_0
    \sqrt{\mathbb{I}_0})\,,
\end{equation}
in place of (\ref{eq:appendix_b2-1}), where
\begin{equation}
    b_{\tau} = \left\{\frac{2m}{(\hbar X_{\tau})^2\, \omega_{\tau}}\,
    (\dot{X}_{\tau}\,\xi_1 + \xi_2)^2 + \frac{2 m \omega_{\tau}}{\hbar^2}\, (\xi_1)^2\right\}^{1/2}\,,\label{eq:appendix_b1-1}
\end{equation}
and $b_0$ is given by (\ref{eq:appendix_b1-1}) with ($\xi_1
\leftrightarrow \xi_2$), ($\omega_0 \leftrightarrow \omega_{\tau}$)
and ($\dot{X}_{\tau} \to Y_{\tau}$). Eqs.
(\ref{eq:wigner-oscillator1-0}) and (\ref{eq:appendix_b1}) give
\begin{equation}\label{eq:appendix_b2}
    \<(\mathbb{I}_{\tau})^0\>_{{\scriptscriptstyle {\mathbb P}}^{(\mbox{\tiny f})}} =
    \frac{m}{(2\pi\hbar^3)\,Z_{\beta}(\omega_0)\,|X_{\tau}|}\,
    \mbox{sech}\left(\frac{\beta\hbar\omega_0}{2}\right)\, \int\int
    d\xi_1\,d\xi_2\; \Lambda_0(\xi_1,\xi_2)\,,
\end{equation}
where
\begin{equation}\label{eq:appendix_b2-1-1-0}
    \Lambda_n(\xi_1,\xi_2) = \lim_{a_{\tau}\to 0} \int_0^{\infty} d\mathbb{I}_{\tau}\;
    (\mathbb{I}_{\tau})^n \times J_0(b_{\tau} \sqrt{\mathbb{I}_{\tau}})\;
    e^{-(a_{\scriptscriptstyle \tau})^2\, \mathbb{I}_{\tau}}
    \int_0^{\infty}
    d\mathbb{I}_0\; J_0(b_0 \sqrt{\mathbb{I}_0})\;
    e^{-(a_{\scriptscriptstyle 0})^2\, \mathbb{I}_{0}}
\end{equation}
with $(a_0)^2 = (2/\hbar)\, \mbox{tanh}(\beta\hbar\omega_0/2)$. We
now apply the identity \cite{ABR65}
\begin{equation}\label{eq:bessel-identity0}
    \int_0^{\infty} dt\; t^{\mu-1}\; \exp(-a^2 t^2)\;
    J_{\nu}(b t) = \frac{(b/2)^{\nu}\; \Gamma((\mu+\nu)/2)}{2\,(a^{\mu+\nu})\; \Gamma(\nu+1)}\; _1F_1\left(\frac{\mu+\nu}{2};\nu+1;\frac{-b^2}{4 a^2}\right)
\end{equation}
with $(\mu=2, \nu=0)$ and $_1F_1(1;1;z) = e^z$ to Eq.
(\ref{eq:appendix_b2-1-1-0}) with $n=0$ such that
\begin{equation}\label{eq:appendix2}
    \Lambda_0(\xi_1,\xi_2) = \lim_{a_{\tau}\to 0} \frac{1}{(a_\tau)^2}\,
    \left.e^{-\gamma\,(b_{\scriptscriptstyle\,\tau})^2\,(2\,a_{\scriptscriptstyle
    \tau})^{-2}}\right|_{\gamma=1}\times
    \frac{1}{(a_0)^2}\, e^{-(b_0)^2\, (2\,a_0)^{-2}}\,.
\end{equation}
The integration of this over $\xi_1$ and $\xi_2$ in Eq.
(\ref{eq:appendix_b2}) will give rise to
\begin{equation}\label{eq:appendix3-0}
    \int\int d\xi_1\,d\xi_2\; \Lambda_0(\xi_1,\xi_2) = \lim_{a_{\scriptscriptstyle \tau}\to 0} \frac{2 \pi \hbar^2/m}{(a_0)^2\, (a_\tau)^2}\,
    \left.F_{\tau}(\gamma)\, G_{\tau}(\gamma)\right|_{\gamma=1} = \frac{2 \pi \hbar^2/m}{(a_0)^2}\, |X_{\tau}|\,,
\end{equation}
where
\begin{eqnarray}\label{eq:appendix4}
    &&F_{\tau}(\gamma) = \left\{\frac{\gamma (\dot{X}_{\tau})^2}{(a_{\tau} X_{\tau})^2\,\omega_{\tau}} + \frac{\gamma \omega_{\tau}}{(a_{\tau})^2} +
    \frac{1}{(a_0 X_{\tau})^2\,\omega_0}\right\}^{-1/2}\;\;\; ;\;\;\; G_{\tau}(\gamma) =\\
    &&\left\{\frac{\gamma}{(a_{\tau} X_{\tau})^2\,\omega_{\tau}} + \frac{\omega_0}{(a_0)^2} +
    \frac{(Y_{\tau})^2}{(a_0 X_{\tau})^2\,\omega_0} -
    [F_{\tau}(\gamma)]^2\, \left(\frac{\gamma \dot{X}_{\tau}}{(a_{\tau} X_{\tau})^2\,\omega_{\tau}} + \frac{Y_{\tau}}{(a_0 X_{\tau})^2\,\omega_0}\right)^2
    \right\}^{-1/2}\,.\n
\end{eqnarray}
Finally, it follows that
$\<(\mathbb{I}_{\tau})^0\>_{{\scriptscriptstyle {\mathbb
P}}^{(\mbox{\tiny f})}} = 1$ indeed.

Then, it is straightforward to obtain the first moment [cf. Eq.
(\ref{eq:appendix_b2})]
\begin{equation}\label{eq:appendix_b2-1-0}
    \<\mathbb{I}_{\tau}\>_{{\scriptscriptstyle {\mathbb P}}^{(\mbox{\tiny f})}} =
    \frac{m}{(2\pi\hbar^3)\,Z_{\beta}(\omega_0)\,|X_{\tau}|}\,
    \mbox{sech}\left(\frac{\beta\hbar\omega_0}{2}\right) \int\int
    d\xi_1\,d\xi_2\; \Lambda_1(\xi_1,\xi_2)\,.
\end{equation}
We now apply (\ref{eq:bessel-identity0}) with $(\mu=4, \nu=0)$ and
$_1F_1\left(2;1;z\right) = (1+z)\,e^z = (1 +
\partial_{\gamma})\,e^{\gamma z}|_{\gamma=1}$ to (\ref{eq:appendix_b2-1-1-0}) with
$n=1$ such that
\begin{equation}\label{eq:appendix2-2}
    \Lambda_1(\xi_1,\xi_2) = \lim_{a_{\tau}\to 0} \frac{1}{(a_\tau)^4} \left.\left(1 +
    \partial_{\gamma}\right)\,
    e^{-\gamma\,(b_{\scriptscriptstyle\,\tau})^2\,(2\,a_{\scriptscriptstyle
    \tau})^{-2}}\right|_{\gamma=1}\times
    \frac{1}{(a_0)^2}\,e^{-(b_0)^2\, (2\,a_0)^{-2}}\,.
\end{equation}
Next, similarly to Eq. (\ref{eq:appendix3-0}), we can obtain
\begin{equation}\label{eq:appendix3}
    \int\int d\xi_1\,d\xi_2\; \Lambda_1(\xi_1,\xi_2) = \lim_{a_{\scriptscriptstyle \tau}\to 0} \frac{2 \pi \hbar^2/m}{(a_0)^2\, (a_\tau)^4} \left.(1 +
    \partial_{\gamma})\, F_{\tau}(\gamma)\,
    G_{\tau}(\gamma)\right|_{\gamma=1}\,.
\end{equation}
This finally simplifies to
$(2\pi\hbar^2)\,|X_{\tau}|\,\mathbb{K}_{\tau}/\{m (a_0)^4\}$ [cf.
(\ref{eq:dimless-1})]. Therefore, Eq. (\ref{eq:appendix_b2-1-0})
reduces to (\ref{eq:phase-space10}). Likewise, we can also evaluate
the first moment $\<\mathbb{I}_0\>_{{\scriptscriptstyle {\mathbb
P}}^{(\mbox{\tiny f})}}$ with the help of $(a_0 \leftrightarrow
a_{\tau})$ and $(b_0 \leftrightarrow b_{\tau})$ in
(\ref{eq:appendix2-2}).

We can further evaluate the higher-order moments in the form of
$\<(\mathbb{I}_{\tau})^n\, (\mathbb{I}_0)^m\>_{{\scriptscriptstyle
{\mathbb P}}^{(\mbox{\tiny f})}}$ by applying the same techniques
with the help of Eqs. (\ref{eq:appendix_b2-1-1-0}),
(\ref{eq:bessel-identity0}) and (\ref{eq:appendix4}) as well as the
recurrence relation $(b-a)\,_1F_1(a-1;b;z) + (2a - b +
z)\,_1F_1(a;b;z) - a\,_1F_1(a+1;b;z)$ \cite{ABR65}; e.g., for the
second moment $\<(\mathbb{I}_{\tau})^2\>_{{\scriptscriptstyle
{\mathbb P}}^{(\mbox{\tiny f})}}$, we have
(\ref{eq:bessel-identity0}) with $(\mu=6, \nu=0)$ and
$_1F_1\left(3;1;z\right) = (z^2 + 4z + 2)\,e^z/2$ [cf. Eqs.
(\ref{eq:higher-order-moments-1})-(\ref{eq:higher-order-moments-3})].
Likewise, the third moments can be evaluated
\begin{subequations}
\begin{eqnarray}
    \<(\mathbb{I}_{\tau})^3\>_{{\scriptscriptstyle {\mathbb P}}^{(\mbox{\tiny f})}} &=&
    \{15\,({\mathbb K}_{\tau})^2 - 9\}\; {\mathbb K}_{\tau}\; \{\<\mathbb{I}_0\>_{{\scriptscriptstyle {\mathbb P}}^{(\mbox{\tiny f})}}\}^3\;\; ;\;\;
    \<(\mathbb{I}_0)^3\>_{{\scriptscriptstyle {\mathbb P}}^{(\mbox{\tiny f})}} = 6\; \{\<\mathbb{I}_0\>_{{\scriptscriptstyle {\mathbb P}}^{(\mbox{\tiny f})}}\}^3\label{eq:3rd-moments-1}\\
    \<(\mathbb{I}_{\tau})^2\,\mathbb{I}_0\>_{{\scriptscriptstyle {\mathbb P}}^{(\mbox{\tiny f})}} &=&
    \{9\,({\mathbb K}_{\tau})^2 - 3\}\; {\mathbb K}_{\tau}\; \{\<\mathbb{I}_0\>_{{\scriptscriptstyle {\mathbb P}}^{(\mbox{\tiny f})}}\}^3\;\; ;\;\;
    \<\mathbb{I}_{\tau}\,(\mathbb{I}_0)^2\>_{{\scriptscriptstyle {\mathbb P}}^{(\mbox{\tiny f})}} =
    6\; {\mathbb K}_{\tau}\; \{\<\mathbb{I}_0\>_{{\scriptscriptstyle {\mathbb P}}^{(\mbox{\tiny f})}}\}^3\,.\label{eq:3rd-moments-2}
\end{eqnarray}
\end{subequations}

Finally, we verify that the internal energy difference as a
picture-independent quantity can be evaluated also in the TPM
framework: To do so, we consider the expectation value
$\<\hat{H}(\omega_{\tau})\>_{\scriptstyle{\rho_{\tau}}} =
\mbox{Tr}\{\hat{H}(\omega_{\tau})\,\hat{\mathcal
U}(\tau)\,\hat{\rho}_0\,\hat{\mathcal U}^{\dagger}(\tau)\}$ with
$\hat{\rho}_0 = \hat{\rho}_{\beta}$. This can be rewritten as
\begin{equation}\label{eq:appendix12}
    \int dy\,dy'\,dx\,dx'\, \left\<y\left|\left(\frac{\hat{p}^2}{2m} + \frac{m\,\omega_{\tau}^2\,\hat{x}^2}{2}\right)\right|x\right\>\; K(x;\tau|x';0)\;
    K^{\ast}(y;\tau|y';0)\; \<x'|\hat{\rho}_{\beta}|y'\>\,,
\end{equation}
where the propagator $K(\cdots)$ in Eq. (\ref{eq:propagator-osc1})
and
\begin{eqnarray}\label{eq:appendix14}
    \<x|\hat{\rho}_{\beta}|y\> &=& \left\{\left(\frac{m\omega_0}{\pi\hbar}\right)\,
    \mbox{tanh}(\beta\hbar\omega_0/2)\right\}^{1/2} \times\\
    && \exp\left[-\frac{m\omega_0}{4\hbar} \left\{(x+y)^2\,
    \mbox{tanh}(\beta\hbar\omega_0/2) + (x-y)^2\,
    \mbox{coth}(\beta\hbar\omega_0/2)\right\}\right]\n
\end{eqnarray}
\cite{TAN07}. Then, it is straightforward to find that
$\<\hat{H}(\omega_{\tau})\>_{\scriptstyle{\rho_{\tau}}} =
\omega_{\tau}\,\<\mathbb{I}_{\tau}\>_{{\scriptscriptstyle {\mathbb
P}}^{(\mbox{\tiny f})}}$.

%%%%%%%%%%%%%%%%%%%%%%%%%%%%%%%%%%%%%%%%%%%%%%%%%%%%%%%%%%%%%%%%%%%%%%%%%%%%%
\section{Evaluations of Eqs. (\ref{eq:phase-space3-3-C0}) and (\ref{eq:phase-space3-3-C})}\label{sec:appendix2}
%%%%%%%%%%%%%%%%%%%%%%%%%%%%%%%%%%%%%%%%%%%%%%%%%%%%%%%%%%%%%%%%%%%%%%%%%%%%%
%
We follow the steps similar to those for the derivation of Eq.
(\ref{eq:appendix_b1}): By employing the identity
\begin{equation}\label{eq:dist9}
    \int_0^{2\pi} d\phi\; e^{i\,\{A\,(\sin\phi) + B\,(\cos\phi)\}}\, (\sin\phi) =
    \frac{\partial_{\scriptscriptstyle A}}{i} \int_0^{2\pi} d\phi\; e^{i\,\{A\,(\sin\phi) + B\,(\cos\phi)\}} =
    -2\pi i\,\partial_{\scriptscriptstyle A}\,J_0(\sqrt{A^2 + B^2})
\end{equation}
with $d J_0(z)/dz = -J_1(z)$, we can finally obtain
\begin{eqnarray}
    && \int_0^{2\pi}\int_0^{2\pi}
    d\phi_{\tau}\, d\phi_0\; \tilde{T}_{\mbox{\tiny{W}}}^{(\mbox{\scriptsize
    f})}(\phi_{\tau},\mathbb{I}_{\tau};\tau|\phi_0,\mathbb{I}_0;0)\; (\sin\phi_0)\n\\
    &=& \frac{-i\,m}{\hbar^2\,|X(\tau)|}
    \int\int d\xi_1\,d\xi_2\; J_0(b_{\tau} \sqrt{\mathbb{I}_{\tau}})\; J_1(b_0
    \sqrt{\mathbb{I}_0})\; (A_0/b_0)\label{eq:dist10-3}
\end{eqnarray}
with $A_0 = \sqrt{2m/\omega_0}\, (\hbar\,|X_{\tau}|)^{-1}\,
(Y_{\tau}\,\xi_2 + \xi_1)$, which is purely imaginary. Therefore,
its real part $C^{(\mbox{\scriptsize
f})}(\mathbb{I}_{\tau}|\mathbb{I}_0)$ becomes zero indeed.

Likewise, we can simplify Eq. (\ref{eq:phase-space3-3-C}) into
\begin{equation}
    D^{(\mbox{\scriptsize f})}(\mathbb{I}_{\tau}|\mathbb{I}_0) = \frac{m}{\hbar^2\,|X_{\tau}|}
    \int\int d\xi_1\,d\xi_2\; J_0(b_{\tau} \sqrt{\mathbb{I}_{\tau}})\; \left\{\frac{J_1(b_0
    \sqrt{\mathbb{I}_0})}{b_0
    \sqrt{\mathbb{I}_0}} - \frac{(A_0)^2\, J_2(b_0
    \sqrt{\mathbb{I}_0})}{(b_0)^2}\right\}\,,
\end{equation}
which is real-valued. Here we also used $d J_1(z)/dz = J_0(z) -
J_1(z)/z$ and $J_0(z) + J_2(z) = 2\,J_1(z)/z$.

\newpage
\begin{figure}[htb]
\centering\hspace*{-2cm}\vspace*{2cm}{
\includegraphics[scale=0.7]{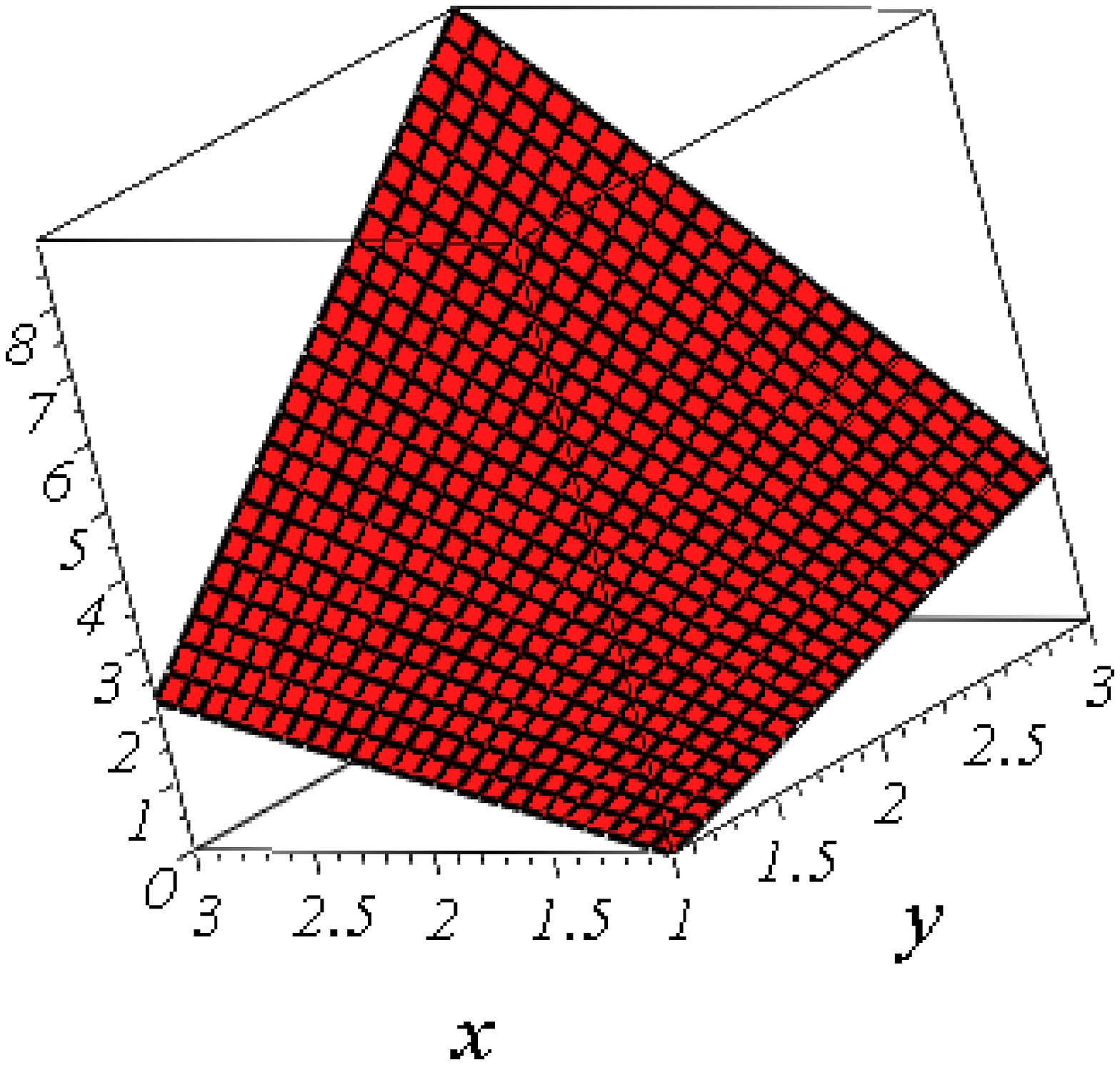}
\caption{\label{fig:fig1}}}
\end{figure}
\begin{figure}[htb]
\centering\hspace*{-2cm}\vspace*{2cm}{
\includegraphics[scale=0.7]{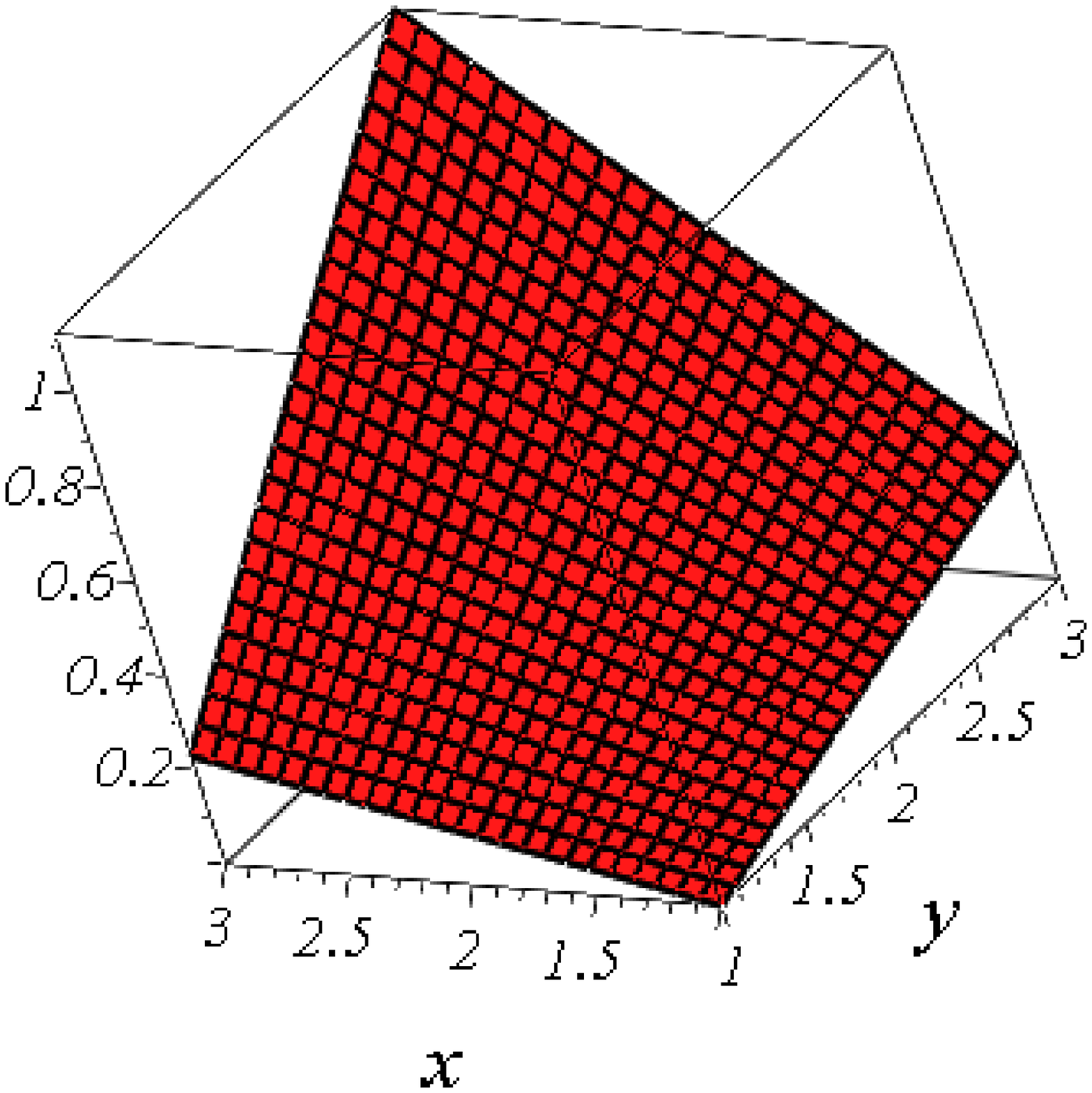}
\caption{\label{fig:fig2}}}
\end{figure}

Fig.~\ref{fig:fig1}: (Color online) The (rescaled) dimensionless
quantity $z_1 = \{\Delta U(\tau) -
\left\<\mathbb{W}\right\>_{{\scriptscriptstyle {\mathbb
P}}^{(\mbox{\tiny f})}}\}/(12 \hbar\omega_0)$ in
(\ref{eq:second-law-first-part1}) versus $(x =
\omega_{\tau}/\omega_0, y = K_{\tau})$. The dimensionless inverse
temperature $ \beta\hbar\omega_0 = 8$ (the low-temperature regime);
cf. Fig. \ref{fig:fig2}.
\newpage
Fig.~\ref{fig:fig2}: (Color online) The (rescaled) dimensionless
quantity $z_2 = \{\Delta U(\tau) -
\left\<\mathbb{W}\right\>_{{\scriptscriptstyle {\mathbb
P}}^{(\mbox{\tiny f})}}\}/(12 \hbar\omega_0)$ in
(\ref{eq:second-law-first-part1}) versus $(x =
\omega_{\tau}/\omega_0, y = K_{\tau})$. The dimensionless inverse
temperature $ \beta\hbar\omega_0 = 4$ (the high-temperature regime).
For comparison with Fig. \ref{fig:fig1}, we see that $z_2$ is
smaller than $z_1$.
\newpage
\begin{figure}[htb]
\centering\hspace*{-2cm}\vspace*{2cm}{
\includegraphics[scale=0.7]{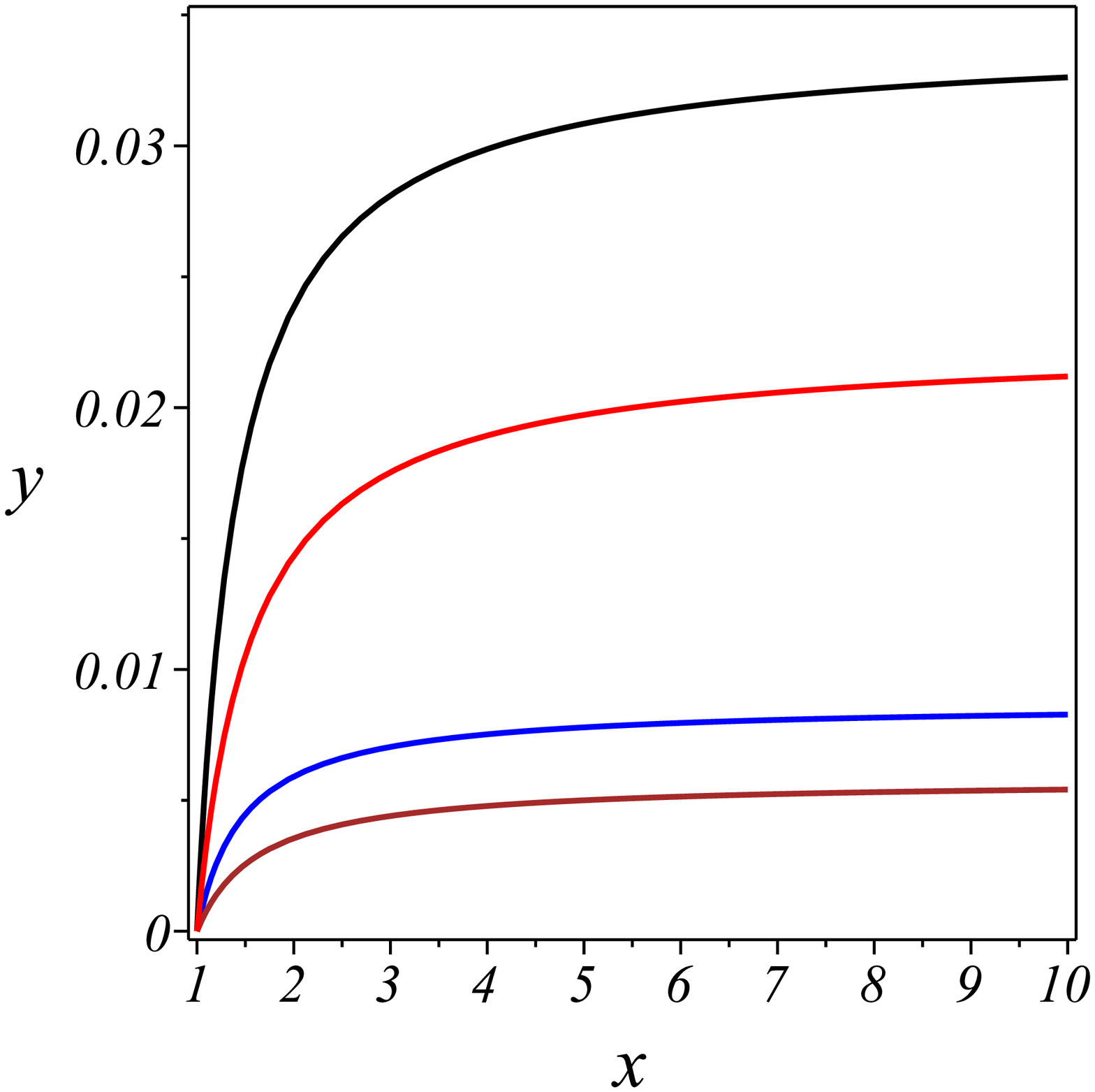}
\caption{\label{fig:fig3}}}
\end{figure}
Fig.~\ref{fig:fig3}: (Color online) The ratio $y =
\mathbb{Q}_{\mbox{\scriptsize d}}^{(\mbox{\scriptsize
f})}/\mathbb{Q}_{\mbox{\scriptsize q}}^{(\mbox{\scriptsize f})}$
after Eq. (\ref{eq:finer-form-0}) versus $x = K_{\tau}$. Let both
dimensionless quantities $\tilde{\beta} = \beta\hbar\omega_0$ and
$\omega = \omega_{\tau}/\omega_0$. From top to bottom: (high
temperature $\tilde{\beta}=5,\omega=2$, black); (high temperature
$\tilde{\beta}=5,\omega=3$, red); (low temperature
$\tilde{\beta}=7,\omega=2$, blue); (low temperature
$\tilde{\beta}=7,\omega=3$, brown). This result consists with the
fact that if $K_{\tau}$ increases (i.e., more non-adiabatic), then
$\mathbb{Q}_{\mbox{\scriptsize d}}^{(\mbox{\scriptsize f})}$
increases. From $y < 1$, it is also noted that the quantum heat
$\mathbb{Q}_{\mbox{\scriptsize q}}^{(\mbox{\scriptsize f})}$ is
large enough (as compared with the dissipative heat
$\mathbb{Q}_{\mbox{\scriptsize d}}^{(\mbox{\scriptsize f})}$) in
such a thermally isolated system, particularly in the
low-temperature regime. Therefore, the quantum heat should be
treated separately without being neglected, as in our framework.
\newpage
\begin{figure}[htb]
\centering\hspace*{-2cm}\vspace*{2cm}{
\includegraphics[scale=0.7]{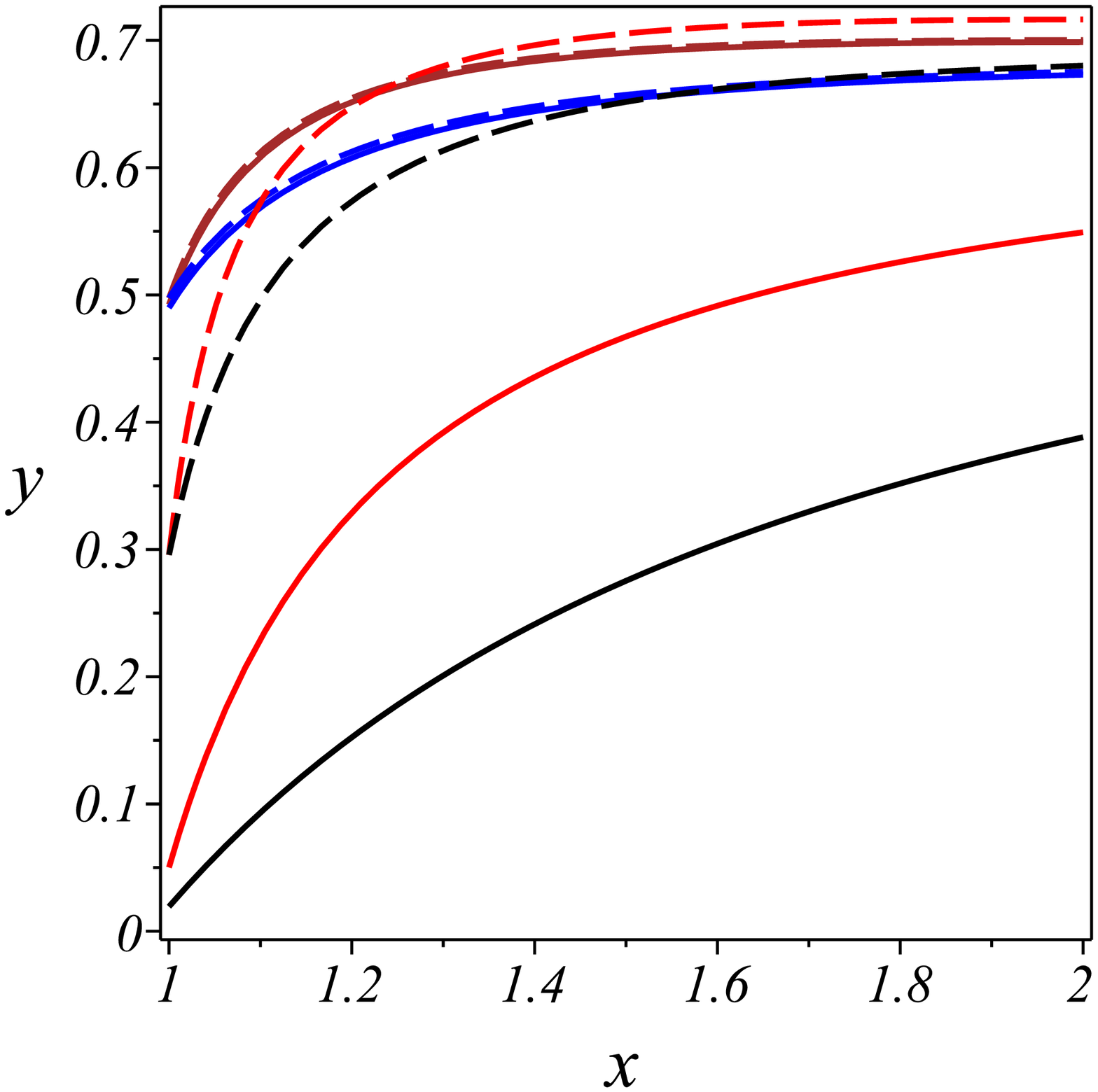}
\caption{\label{fig:fig4}}}
\end{figure}
Fig.~\ref{fig:fig4}: (Color online) Solid: The relative variance
$y_1 = \<(\widetilde{\Delta \mathbb{W}})^2\>_{{\scriptscriptstyle
{\mathbb P}}^{(\mbox{\tiny f})}}$ in Eq. (\ref{eq:variance-work1})
versus $x = K_{\tau}$. Let both dimensionless quantities
$\tilde{\beta} = \beta\hbar\omega_0$ and $\omega =
\omega_{\tau}/\omega_0$. From top to bottom (at $x=1.01$): (high
temperature $\tilde{\beta}=0.2,\omega=2$, brown); (high temperature
$\tilde{\beta}=0.2,\omega=3$, blue); (low temperature
$\tilde{\beta}=2,\omega=2$, red); (low temperature
$\tilde{\beta}=2,\omega=3$, black). Dash: The relative variance $y_2
= \<(\widetilde{\Delta {\mathrm
w}})^2\>_{\scriptscriptstyle{P^{(\mbox{\tiny
f})}}}/(\hbar\omega_0)^2$ in Eq.
(\ref{eq:difference-bet-tpm-and-mine1}). From top to bottom: The
same as for $y_1$. We see that $y_1$ and $y_2$ are almost identical
in the high-temperature regime ($\tilde{\beta} = 0.2$).
\newpage
\begin{figure}[htb]
\centering\hspace*{-2cm}\vspace*{2cm}{
\includegraphics[scale=0.7]{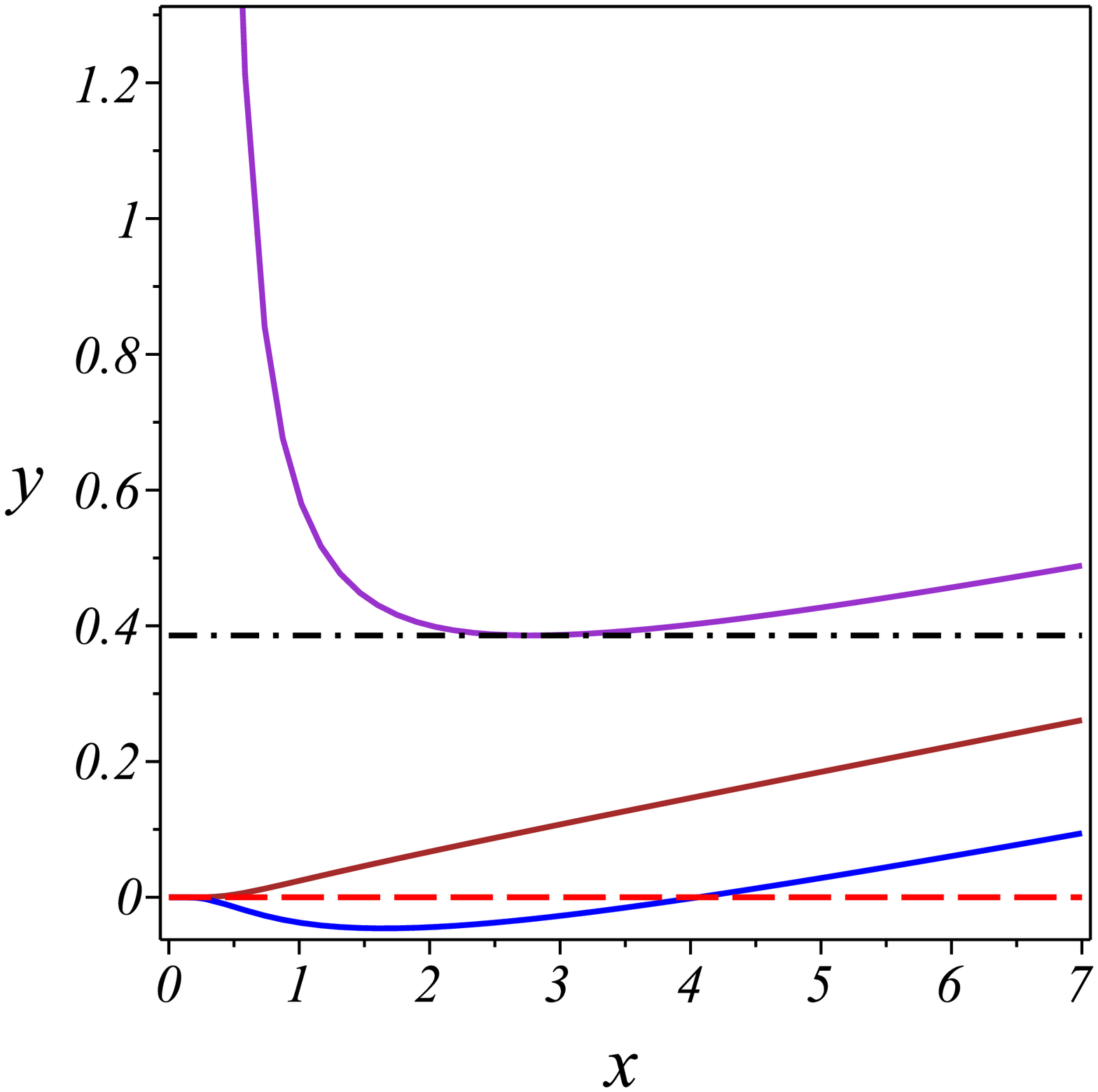}
\caption{\label{fig:fig5}}}
\end{figure}
Fig.~\ref{fig:fig5}: (Color online) The dimensionless ``dissipative
heat'' $y =
\mathbb{Q}_{\scriptscriptstyle{\Upsilon},\mbox{\scriptsize{d}}}^{(\mbox{\scriptsize
f})}/\hbar\omega_0$ in Eq. (\ref{eq:dissipative-heat-0-huimi})
versus the dimensionless temperature $x =
(\beta\hbar\omega_0)^{-1}$. Let $\omega_{\tau}/\omega_0 = 1.3$. From
top to bottom (solid): ($\Upsilon = P$, purple); ($\Upsilon = W$,
brown); ($\Upsilon = Q$, blue). For comparison, $y=0$ (dash, red).
We see that the $P$-curve shows its minimum value $y = 0.386$
(dashdot, black) at $x = 2.776$, which is physically inconsistent,
and the $Q$-curve can be negative valued. The $y$-values of all
three curves increase with the temperature in the high-temperature
regime. Therefore, it is the $W$-curve only that consists with the
thermodynamics.
\newpage
\begin{figure}[htb]
\centering\hspace*{-2cm}\vspace*{2cm}{
\includegraphics[scale=0.7]{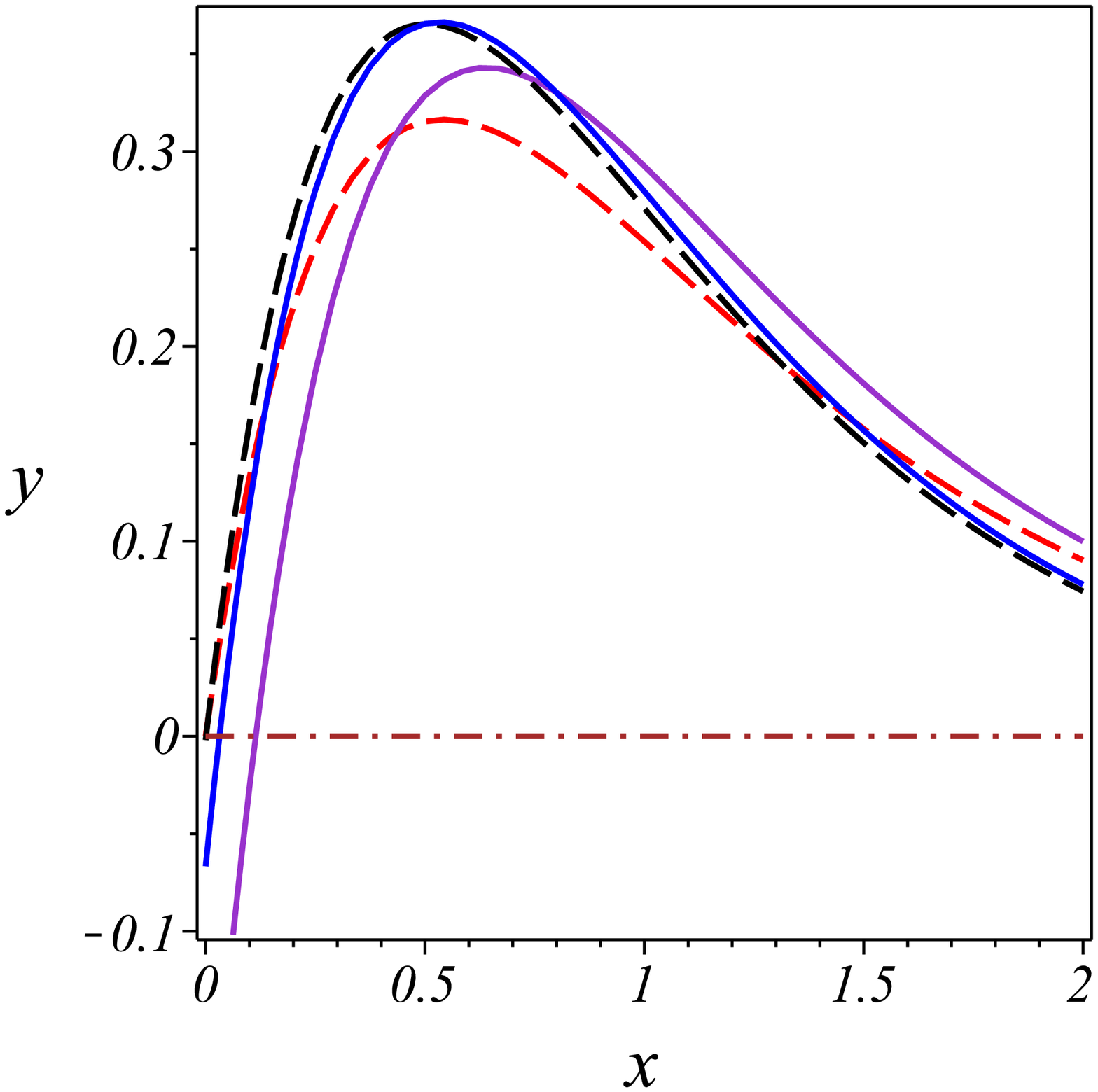}
\caption{\label{fig:fig6}}}
\end{figure}
Fig.~\ref{fig:fig6}: (Color online) The dimensionless Wigner
function $y = \pi\hbar\,\{(1-\gamma)\,W_{\beta}(\mathbb{I}) +
\gamma\,W_n(\mathbb{I})\}$ with $n=1$ after Eq.
(\ref{eq:phase-space7-sig}) versus the dimensionless action
$x=\mathbb{I}/\hbar$. Let the dimensionless inverse temperature
$\tilde{\beta} = \beta\hbar\omega_0$. Dash: From top to bottom at
$x=0.5$, ($\tilde{\beta} = 2.5$, black) with $\gamma = 0.460$, which
is its threshold value $(\gamma_{\mbox{\scriptsize
th},\scriptscriptstyle{\underline{1}}})_{\beta}$; ($\tilde{\beta} =
1$, red) with $\gamma = 0.316 = (\gamma_{\mbox{\scriptsize
th},\scriptscriptstyle{\underline{1}}})_{\beta}$. Solid: From top to
bottom at $x=0.5$, ($\tilde{\beta} = 2.5$, blue) with $\gamma =
0.495
> (\gamma_{\mbox{\scriptsize
th},\scriptscriptstyle{\underline{1}}})_{\beta}$ thus showing its
negativity; ($\tilde{\beta} = 1$, purple) with $\gamma = 0.490
> (\gamma_{\mbox{\scriptsize
th},\scriptscriptstyle{\underline{1}}})_{\beta}$ thus showing its
negativity. For comparison, $y = 0$ (dashdot, brown).
\newpage
\begin{figure}[htb]
\centering\hspace*{-2cm}\vspace*{2cm}{
\includegraphics[scale=0.7]{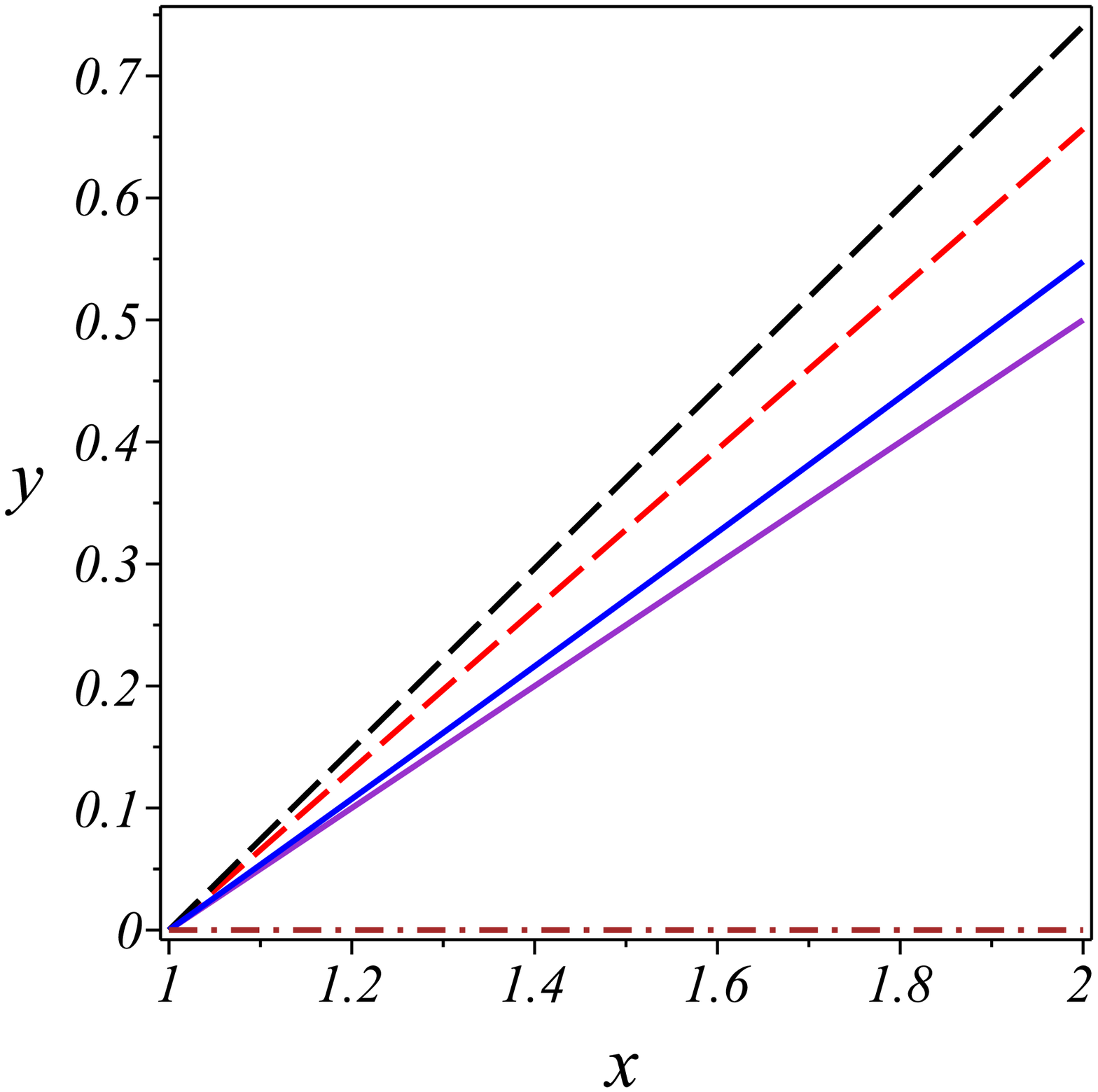}
\caption{\label{fig:fig7}}}
\end{figure}
Fig.~\ref{fig:fig7}: (Color online) A periodic external driving with
$\omega_{\tau} = \omega_0$ and the partially thermal initial state
with $\hat{\sigma}_1 = |1\>\<1|$. Let the dimensionless inverse
temperature $\beta\hbar\omega_0 = 2$. The dimensionless internal
energy difference $y_1 = \Delta
U_{\underline{1}}(\tau)/\hbar\omega_0$ in (\ref{eq:dist15-0}) versus
$x = K_{\tau}$ (dash). From top to bottom: ($\gamma = 0.1$, black);
($\gamma = 0$, red). The dimensionless average work $y_2 =
\left\<\mathbb{W}\right\>_{{\scriptscriptstyle {\mathbb
P}}_{\underline{1}}^{(\mbox{\tiny f})}}/\hbar\omega_0$ (solid). From
top to bottom: ($\gamma = 0.1$, blue); ($\gamma = 0$, purple). The
dimensionless free energy difference $y_3 = \Delta
F_{\beta}^{(\mbox{\scriptsize f})}/\hbar\omega_0 = 0$ (dashdot,
brown). For the evaluation of the curve $y_2$, we used $\ln(1+z)
\approx z$ for $|z| \ll 1$ (i.e., $\gamma = 0.1 \ll 1$).
\newpage
\begin{figure}[htb]
\centering\hspace*{-2cm}\vspace*{2cm}{
\includegraphics[scale=0.63]{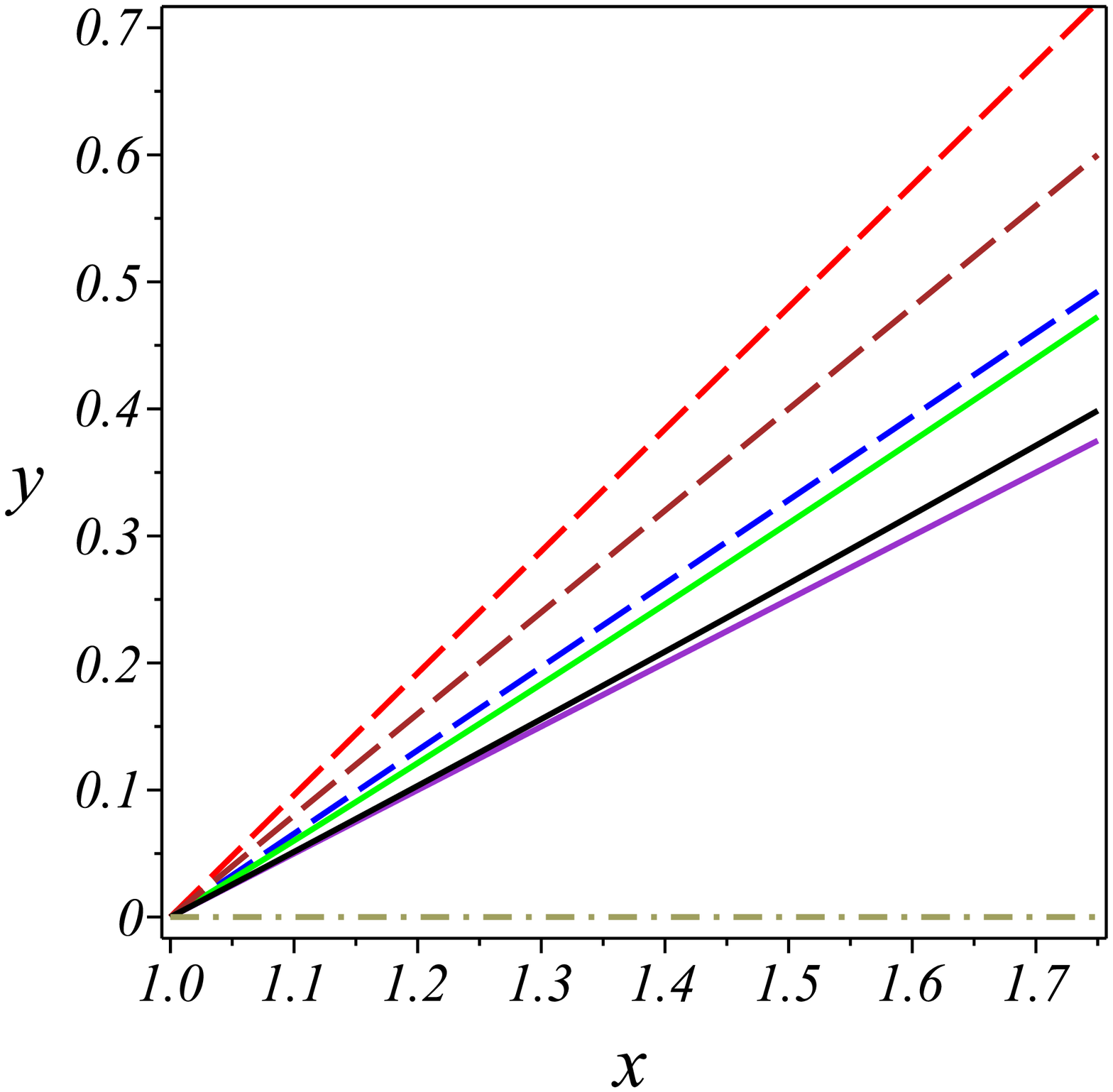}
\caption{\label{fig:fig8}}}
\end{figure}
Fig.~\ref{fig:fig8}: (Color online) A periodic external driving with
$\omega_{\tau} = \omega_0$ and the partially thermal initial state
with $\hat{\sigma}_3 = (|0\> + |1\> + |2\>)(\<0| + \<1| + \<2|)/3$.
Let the dimensionless inverse temperature $\beta\hbar\omega_0 = 2$.
The dimensionless internal energy difference $y_1 = \Delta
U_{\underline{3}}(\tau)/\hbar\omega_0$ in Eq.
(\ref{eq:internal-energy_diff12}) versus $x = K_{\tau}$ (dash). From
top to bottom: ($\gamma = 0.17$, red) with coherence; ($\gamma =
0.17$, brown) without coherence, i.e., $y_1 \to \Delta
U_{\utilde{3}}(\tau)/\hbar\omega_0$ in Eq.
(\ref{eq:internal-energy_diff12-0}); ($\gamma = 0$, blue). The
dimensionless average work $y_2 =
\left\<\mathbb{W}\right\>_{{\scriptscriptstyle {\mathbb
P}}_{\underline{3}}^{(\mbox{\tiny f})}}/\hbar\omega_0$ after Eq.
(\ref{eq:work-dist-new_1-sig-11-1}) (solid). From top to bottom:
($\gamma = 0.17$, green) with coherence; ($\gamma = 0.17$, black)
without coherence, i.e., $y_2 \to
\left\<\mathbb{W}\right\>_{{\scriptscriptstyle {\mathbb
P}}_{\utilde{3}}^{(\mbox{\tiny f})}}/\hbar\omega_0$; ($r=0$,
purple). The dimensionless free energy difference $y_3 = \Delta
F_{\beta}^{(\mbox{\scriptsize f})}/\hbar\omega_0 = 0$ (dashdot,
khaki). For the evaluation of the curve $y_2$, we used $\ln(1+z)
\approx z$ for $|z| \ll 1$ (i.e., $\gamma = 0.17 \ll 1$).
\end{document}